\documentclass{statsoc}
\usepackage[a4paper]{geometry}
\usepackage[T1]{fontenc}
\usepackage[utf8]{inputenc}
\usepackage{graphicx}
\usepackage{pgf,tikz}
\usepackage{pdfpages}

\usepackage{amsfonts, amsmath, amssymb, amsthm, bbm}
\usepackage{MnSymbol}
\usepackage[mathscr]{euscript}
\usepackage{float}
\usepackage{multirow}
\usepackage{natbib}

\usepackage{color}
\definecolor{darkred}{RGB}{100,0,0}
\definecolor{darkgreen}{RGB}{0,100,0}
\definecolor{darkblue}{RGB}{0,0,150}

\usepackage{hyperref}
\hypersetup{colorlinks=true, linkcolor=darkred, citecolor=darkgreen, urlcolor=darkblue}

\usepackage{mathtools}
\mathtoolsset{showonlyrefs}

\usepackage{etoolbox}

\makeatletter
\patchcmd{\@makecaption}
  {\parbox}
  {\advance\@tempdima-\fontdimen2} 
  {}{}
\makeatother 

\usepackage[ruled,vlined]{algorithm2e}

\newtheorem{thm}{Theorem}

\newtheorem{lem}{Lemma}
\newtheorem{cor}{Corollary}
\newtheorem{assump}{Assumption}

\theoremstyle{remark}

\newtheorem{rem}{Remark}

\def\beq{\begin{equation}} 
\def\eeq{\end{equation}}
\def\beqn{\begin{eqnarray*}}
\def\eeqn{\end{eqnarray*}}
\def\Bitem{\begin{itemize}\setlength{\itemsep}{.2in}}
\def\bitem{\begin{itemize}\setlength{\itemsep}{.05in}}
\def\eitem{\end{itemize}}
\def\Benum{\begin{enumerate}\setlength{\itemsep}{.2in}}
\def\benum{\begin{enumerate}\setlength{\itemsep}{.05in}}
\def\eenum{\end{enumerate}}
\def\bmult{\begin{multline*}}
\def\emult{\end{multline*}}
\def\bcenter{\begin{center}}
\def\ecenter{\end{center}}
\def\bframe{\begin{frame}}
\def\eframe{\end{frame}}

\def\cA{\mathcal{A}}

\def\cO{\mathcal{O}}

\def\cT{\mathcal{T}}

\newcommand{\E}{\operatorname{\mathbb{E}}}
\renewcommand{\P}{\operatorname{\mathbb{P}}}

\def\eps{\varepsilon}

\def\1{\mathbbm{1}}

\usetikzlibrary{arrows,shapes.arrows,shapes.geometric,shapes.multipart, decorations.pathmorphing,positioning,shapes.swigs}

\title[Proximal Causal Inference for Complex Longitudinal Studies]{Proximal Causal Inference for Complex Longitudinal Studies}

\author{Andrew Ying}
\address{Department of Statistics and Data Science, The Wharton School, University of Pennsylvania,
Philadelphia,
U.S.A.}
\author{Wang Miao}
\address{Department of Probability and Statistics, Peking University,
Beijing,
P.R.C.}
\author{Xu Shi}
\address{Department of Biostatistics, University of Michigan,
Ann Arbor,
U.S.A.}

\author[A. Ying, W. Miao, X. Shi and E.J. Tchetgen Tchetgen]{and Eric J. Tchetgen Tchetgen}
\address{Department of Statistics and Data Science, The Wharton School, University of Pennsylvania,
Philadelphia,
U.S.A.}

\date{}

\begin{document}

\maketitle

\begin{abstract}
A standard assumption for causal inference about the joint effects of time-varying treatment is that one has measured sufficient covariates to ensure that within covariate strata, subjects are exchangeable across observed treatment values, also known as ``sequential randomization assumption (SRA)''. SRA is often criticized as it requires one to accurately measure all confounders. Realistically, measured covariates can rarely capture all confounders with certainty. Often covariate measurements are at best proxies of confounders, thus invalidating inferences under SRA. In this paper, we extend the proximal causal inference (PCI) framework of \cite{miao2018identifying} to the longitudinal setting under a semiparametric marginal structural mean model (MSMM). PCI offers an opportunity to learn about joint causal effects in settings where SRA based on measured time-varying covariates fails, by formally accounting for the covariate measurements as imperfect proxies of underlying confounding mechanisms. We establish nonparametric identification with a pair of time-varying proxies and provide a corresponding characterization of regular and asymptotically linear estimators of the parameter indexing the MSMM, including a rich class of doubly robust estimators, and establish the corresponding semiparametric efficiency bound for the MSMM. Extensive simulation studies and a data application illustrate the finite sample behavior of proposed methods.

\end{abstract}


\section{Introduction}
A common assumption for causal inference from observational longitudinal data is the so-called ``sequential randomization assumption (SRA)'' \citep{robins1986new, robins1987graphical, robins1997causal, robins1997marginal, robins1999association}, which states that at each follow-up time, one has measured a sufficiently rich set of covariates to ensure that conditional on covariate and treatment history, subjects are exchangeable across observed treatment values received at that time point. This fundamental assumption is inherently untestable empirically, without introducing a different untestable assumption, and therefore must be taken on faith even with substantial subject matter knowledge at hand. For this reason, SRA is often the subject of much skepticism, mainly because it hinges on an assumed ability of the investigator to accurately measure covariates relevant to the various confounding mechanisms potentially present in the observational study. Realistically, confounding mechanisms can rarely if ever, be learned with certainty from measured covariates. Therefore, practically in a given observational study, one can at most hope that covariate measurements are proxies of the true underlying confounding mechanism. Such acknowledgement invalidates any causal claim made on the basis of SRA. 

\subsection{Proximal Causal Inference Framework}\label{sec:proximalintro}
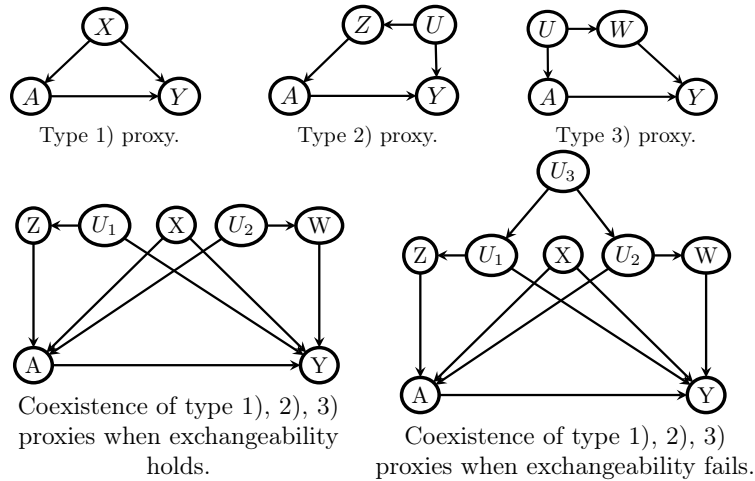
\begin{figure}[H]
\centering
\resizebox{.7\textwidth}{!}{
	\begin{minipage}[b]{0.3\linewidth}
		\centering
		\resizebox{3.5cm}{!}{


\begin{tikzpicture}
\tikzset{line width=1.5pt, outer sep=0pt,
ell/.style={draw,fill=white, inner sep=2pt,
line width=1.5pt},
swig vsplit={gap=5pt, line color right=red,
inner line width right=0.5pt}};

\node[name=X, ell, shape=ellipse]{$X$};

\node[name=A, below left=10mm of X, ell, shape=ellipse]{$A$};

\node[name=Y, below right=10mm of X, ell, shape=ellipse]{$Y$};

\draw[->,line width=1.0pt,>=stealth](X) to (Y);
\draw[->,line width=1.0pt,>=stealth](X) to (A);

\draw[->,line width=1.0pt,>=stealth](A) to (Y);

\end{tikzpicture}

}
		\par Type 1) proxy.
	\end{minipage}
	\begin{minipage}[b]{0.3\linewidth}
		\centering
		\resizebox{3.5cm}{!}{


\begin{tikzpicture}
\tikzset{line width=1.5pt, outer sep=0pt,
ell/.style={draw,fill=white, inner sep=2pt,
line width=1.5pt},
swig vsplit={gap=5pt, line color right=red,
inner line width right=0.5pt}};

\node[name=Z, ell, shape=ellipse]{$Z$};

\node[name=U, right=5mm of Z, ell, shape=ellipse]{$U$};

\node[name=A, below left=10mm of Z, ell, shape=ellipse]{$A$};

\node[name=Y, below right=10mm of Z, ell, shape=ellipse]{$Y$};

\draw[->,line width=1.0pt,>=stealth](U) to (Z);
\draw[->,line width=1.0pt,>=stealth](U) to (Y);
\draw[->,line width=1.0pt,>=stealth](Z) to (A);

\draw[->,line width=1.0pt,>=stealth](A) to (Y);

\end{tikzpicture}

}
        \par Type 2) proxy.
	\end{minipage}
	\begin{minipage}[b]{0.3\linewidth}
		\centering
		\resizebox{3.5cm}{!}{


\begin{tikzpicture}
\tikzset{line width=1.5pt, outer sep=0pt,
ell/.style={draw,fill=white, inner sep=2pt,
line width=1.5pt},
swig vsplit={gap=5pt, line color right=red,
inner line width right=0.5pt}};

\node[name=W, ell, shape=ellipse]{$W$};

\node[name=U, left=5mm of W, ell, shape=ellipse]{$U$};

\node[name=A, below left=10mm of W, ell, shape=ellipse]{$A$};

\node[name=Y, below right=10mm of W, ell, shape=ellipse]{$Y$};

\draw[->,line width=1.0pt,>=stealth](W) to (Y);
\draw[->,line width=1.0pt,>=stealth](U) to (A);
\draw[->,line width=1.0pt,>=stealth](U) to (W);

\draw[->,line width=1.0pt,>=stealth](A) to (Y);

\end{tikzpicture}

}
		\par Type 3) proxy.
	\end{minipage}
	}
	\centering
	\resizebox{.7\textwidth}{!}{
	\begin{minipage}[b]{0.4\linewidth}
		\centering
		\includegraphics[scale=1]{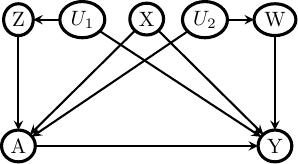}
		\par Coexistence of type 1), 2), 3) proxies when exchangeability holds.
	\end{minipage}\hfill
	\begin{minipage}[b]{0.4\linewidth}
		\centering
		\includegraphics[scale=1]{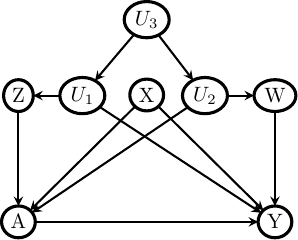}
		\par Coexistence of type 1), 2), 3) proxies when exchangeability fails.
	\end{minipage}}
	\caption{Directed Acyclic Graphs illustrating treatment- and outcome-inducing proxies in a simple point exposure case, where we temporarily term $A$ as the treatment, $Y$ as the outcome, $X$ as the type 1) proxy, $Z$ as the type 2) proxy, $W$ as the type 3) proxy, $U$, $U_1$, $U_2$, $U_3$ as some unmeasured confounders.}
	\label{fig:dag1}
\end{figure}

Instead of relying on SRA on the basis of measured covariates, proximal causal inference essentially requires that the analyst has measured covariates, that can be classified into three bucket types: 1) variables that may be common causes of the treatment and outcome variables; 2) potential treatment-inducing confounding proxies; and 3) potential outcome-inducing confounding proxies. A proxy of type 2) is a potential cause of the treatment which is related with the outcome only through an unmeasured common cause for which the variable is a proxy; while a proxy of type 3) is a potential cause of the outcome which is related with the treatment only through an unmeasured common cause for which the variable is a proxy. Proxies that are neither causes of treatment or outcome variables can belong to either bucket type 2) or 3). An illustration of proxies of types 1) - 3) is given for a simple point exposure case in Figure \ref{fig:dag1}. 

Negative control treatment and outcome variables form a prominent class of proxies that has in recent years received growing interest; e.g. see \cite{lipsitch2010negative, kuroki2014measurement, miao2018identifying, sofer2016negative, shi2020multiply}; \cite{shi2020selective}. With the exception of \cite{tchetgen2020introduction}, prior literature on proxies has largely focused on point exposure studies and not considered joint effects of longitudinal treatments. The current paper builds on initial results obtained in \cite{tchetgen2020introduction} and like the latter, departs from the current practice of assuming that  sequential randomization can be attained upon adjusting for measured time-varying covariates. Instead, we leverage an investigator's ability to classify measured time varying covariates as proxies of types 1), 2) or 3) of  unmeasured time-varying factors that would in principle suffice to account for time-varying confounding. This condition is formalized using the potential outcomes framework in Section \ref{sec:pretwo}. 

Here we briefly introduce the data application we will later analyze, which we use throughout as a running example; additional examples of longitudinal proxies are discussed in supplementary material, also see \cite{tchetgen2020introduction}. In this paper, we aim to evaluate the joint causal effects of the disease-modifying anti-rheumatic therapy Methotrexate (MTX) over time among patients with rheumatoid arthritis (RA). The outcome is the average number of tender joints at end of follow-up, a well-established measure of disease progression.  Although prior studies have established the effectiveness of MTX against premature mortality in RA patients \citep{choi2002methotrexate}, prior analyses relied on SRA and therefore may be susceptible to bias due to residual confounding of time-varying use of MTX by a patient's evolving health status and her potential health-seeking behavior. Fortunately, the available data include important covariates known to be associated with both MTX uptake and disease progression, including demographic, clinical, laboratory, other medication use, and self-reported health status updated over time. Among measured covariates, RA activity measures such as health assessment questionnaire, number of tender joints, patient’s global assessment, and erythrocyte sedimentation rate stand out as likely subject to reporting or measurement error and therefore may be viewed as good candidate proxies of a patient's underlying state of recent disease progression at the source of confounding. Such proxies seldom constitute a common cause of both treatment allocation (MTX use) and disease progression (increased number of tender joints), but may be strongly associated with both treatment and outcome variables to the extent that they share an unmeasured common cause corresponding to the patient's underlying health status and her inherent health-seeking behavior (e.g. health assessment questionnaire, being a measure for quality of life, is a proxy of underlying health status). Thus, such variables provide a candidate set of proxies of type 2) and 3) as we argue throughout the paper and leverage in Section \ref{sec:real} in order to obtain more credible causal estimates of the joint effects of MTX on disease progression.

\subsection{Related Literature and Our Contributions}
As mentioned in the previous section, the proximal causal inference framework is closely related to recent literature on the use of negative control variables to identify and sometimes mitigate confounding bias in the analysis of observational data, see \citet{lipsitch2010negative, kuroki2014measurement, miao2018identifying, shi2020multiply}; \citet{shi2020selective}. Initial results on point identification of causal effects leveraging negative control variables relied on fairly restrictive assumptions such as linear models for the outcome and unmeasured confounding variables \citep{flanders2011method, gagnon2012using, flanders2017new, wang2017confounder}, rank preservation \citep{tchetgen2014control}, monotonicity \citep{sofer2016negative}, or categorical unmeasured confounders \citep{shi2020multiply}. \cite{miao2018identifying} were first to establish sufficient conditions for nonparametric identification of causal effects using a pair of proxies (including negative control variables) in the point treatment setting.

Building upon \cite{miao2018identifying}, recently \cite{tchetgen2020introduction} introduced a potential outcome framework for proximal causal inference, which offers an opportunity to learn about causal effects in point treatment or time-varying treatment settings where the assumption of no unmeasured confounding or sequential randomization on the basis of measured covariates fails. In their work, identification hinges on a longitudinal generalization of \cite{miao2018identifying}, which relies on an assumption that certain Fredholm integral equations of the first kind involving the observed outcome process, admit a solution. For estimation and inference, \cite{tchetgen2020introduction} focused primarily on so-called proximal g-computation, a generalization of Robins’ g-computation algorithm which may be viewed essentially as a maximum likelihood estimator, requiring a correctly specified model restricting the observed data joint distribution. Notably, they propose a proximal recursive two-stage least squares algorithm for point and time-varying treatments. The algorithm remains consistent provided a key linear model restricting the observed data distribution for the outcome holds, even if a linear model restricting the distribution of the time-varying proxies is incorrect. However, recursive two-stage least squares fails to be consistent if the linear outcome model is misspecified. In the point treatment case, \cite{cui2020semiparametric} proposed an alternative set of conditions for nonparametric proximal identification of the average treatment effect and effect on the treated under the assumption that a certain Fredholm integral equation of the first kind involving the treatment data generating mechanism admits a solution. They also developed semiparametric theory for proximal estimation of the average treatment effect (and the treatment effect for the treated), including efficiency bounds for key semiparametric models of interest and characterized proximal doubly robust and locally efficient estimators of the average treatment effect. \cite{deaner2018panel} proposed identification results of the so-called ``conditional average structural function'' (CASF), thus independently establishing identification conditions for the effect of treatment on the treated in the case of point exposure. For longitudinal data, \cite{deaner2018panel} leveraged a Markov condition that lagged treatments have a null causal effect on the outcome. Importantly, unlike \cite{tchetgen2020introduction} who avoid the assumption that past treatments do not have a direct effect on future outcomes, Deaner's Markov conditions essentially reduce a potentially complex longitudinal study involving time-varying treatments, into a series of point exposure studies with past treatment and outcome variables providing a rich source of potential proxies. Estimation in \cite{deaner2018panel} is performed using a penalized sieve minimum distance estimator for the outcome process. \cite{tennenholtz2020off} recently investigated proximal identification in off-policy evaluation for time series, where similar to \cite{deaner2018panel} they leverage Markov restrictions to generate proxies.

In this paper, we aim to develop proximal causal inference and semiparametric theory for complex longitudinal studies when SRA fails to hold due to unmeasured time-varying confounding. Notably, as mentioned above, similar to \cite{tchetgen2020introduction}, we do not impose the Markov condition of \cite{deaner2018panel, tennenholtz2020off} on the effect of lagged treatments, and thus, we allow time-varying confounders (both measured and unmeasured) to mediate the causal effects of past treatment, a widely recognized challenge of complex longitudinal studies routinely encountered in health and social sciences.  In this vein, we aim to make inferences about the parameters indexing a marginal structural mean model (MSMM) \citep{robins1997marginal, robins1999association, robins2000marginal, robins2000marginalb}, a well-established class of counterfactual models for the joint causal effects of time-varying treatments subject to time-varying confounding. In recent work, \cite{tchetgen2020introduction} gave sufficient conditions for proximal nonparametric identification of the joint effects of time-varying treatments, therefore establishing that one could in principle, sample size permitting, estimate saturated MSMs using the proposed proximal framework, simultaneously accounting for measured and unmeasured time-varying confounding. Their results which we briefly review in the next sections is based on so-called {\it outcome confounding bridge functions}, a natural extension of Robins' foundational g-formula to the proximal framework. In this paper, we further this line of work by proposing an alternative identification result via the so-called {\it treatment confounding bridge functions}, a longitudinal generalization of an approach proposed in \cite{cui2020semiparametric} in the point treatment setting. A major contribution of the paper is to provide a general semiparametric theory for proximal inference about MSMM parameters in longitudinal settings leveraging time-varying proxies. Specifically, we derive a rich class of estimators including proximal outcome regression (POR) estimators, proximal inverse probability weighted (PIPW) estimators, and proximal doubly robust (PDR) estimators, which extend existing OR estimators, IPW estimators, and DR estimators derived under SRA \citep{robins1997marginal} and generalize \cite{cui2020semiparametric}'s semiparametric estimators to the longitudinal setting. Furthermore, we establish the semiparametric efficiency bound for the parameters of an MSMM assuming the observed data distribution is otherwise unrestricted, and we provide a one-step update estimator which is locally efficient in the sense that it attains the efficiency bound at the intersection submodel where all posited models are correctly specified. We emphasize that the contributions made in this paper are non-trivial developments that extend the proximal causal inference framework to one of the most challenging settings encountered in epidemiology and related sciences: complex longitudinal studies with time-varying treatment and, both measured and unmeasured time-varying confounders potentially affected by prior treatment.

Thus, our paper contributes to the growing literature on the identification and inference of MSMM parameters under endogeneity. Recently, \cite{tchetgen2018marginal} developed an instrumental variables approach to identify and estimate MSM parameters without SRA \citep{cui2020semiparametric2, michael2020instrumental}. A key assumption in their work entails an ``independent compliance type'', which rules out any additive interaction between instrument and unmeasured confounders in a longitudinal model for the treatment process. No such restriction is needed in the proximal causal framework.

The remainder of the article is organized as follows. We introduce notation and key assumptions in Section \ref{sec:pretwo}. We provide proximal identification results in Section \ref{sec:identwo}. In Section \ref{sec:semitwo}, we derive the set of influence functions under a semiparametric MSMM. Furthermore, we derive the efficient influence function and thus the semiparametric efficiency bound for the MSMM parameters. In Section \ref{sec:esttwo}, we propose three practical classes of estimators including a rich class of doubly robust estimators. We further apply our proposed estimators to the data application evaluating the joint causal effects of anti-rheumatic therapy Methotrexate (MTX) use over time among patients with rheumatoid arthritis in Section \ref{sec:real}. We end the paper with a discussion in Section \ref{sec:dis}. Proofs, additional regularity conditions, additional theoretical results, and extensive simulations are provided in the supplementary material.

\section{Preliminaries}\label{sec:pretwo}
\subsection{Notation}
For the sake of clarity in the exposition, we restrict the presentation of all results to the two-occasion longitudinal case, and we relegate results and proofs for the general case of arbitrary length of follow-up to the supplementary material. Importantly, this simplification is without loss of generality as the two-occasion case captures the essential complexities of the general case. In this vein, suppose that one has observed $n$ i.i.d. copies of longitudinal data $(Y, \overline A(1), \overline L(1))$, where $Y$ is a measure of an outcome at end of follow-up, $\overline A(1) = (A(0), A(1))$ represents a binary treatment process up to time $1$ and $\overline L(1)$ are observed covariates up to time $1$. We aim to investigate the joint effects of $\overline A(1)$ on the outcome $Y$ through an MSMM. Let $\cA = \{(0,0),(0,1),(1,0),(1,1)\}$ denote the set of possible treatment allocations and $Y_{\overline a(1)}$, $\overline a(1) \in \cA$ denote the potential outcome \citep{robins1986new, robins1987graphical} that would be observed if the treatment process were, possibly contrary to fact, set to $\overline a(1)$. We make the following standard consistency assumption that $Y = Y_{\overline A(1)}$ almost surely, which links observed outcomes and potential outcomes via the observed treatment process. 

The sequential randomization assumption of \cite{robins1986new, robins1987graphical, robins1997causal} is  expressed as, $Y_{\overline a(1)} \perp A(0) |L(0)$ and $Y_{\overline a(1)} \perp A(1) |A(0), \overline L(1)$, which essentially requires that $L(0)$ includes all common causes of $Y$ and $A(0)$, and $A(0), \overline L(1)$ include all common causes of $Y$ and $A(1)$. It is well known that under consistency, SRA and the positivity assumption, the counterfactual mean $\E(Y_{\overline a(1)})$ is nonparametrically identified from the observed data distribution by the g-formula of \cite{robins1986new} $\E(Y_{\overline a(1)}) = \E\{\E[\E(Y|\overline a(1), \overline L(1))|a(0), L(0)]\}$.  Below, we discuss the proximal causal inference framework which offers an alternative set of identification conditions that allows for nonparametric identification of the counterfactual mean $\E(Y_{\overline a(1)})$, even when SRA fails to hold due to possible time-varying unmeasured confounding.

To formally introduce the longitudinal proximal causal inference framework, analogous to \cite{tchetgen2020introduction}, suppose that the observed covariates $\overline L(1)$ consists of three  types $(\overline X(1), \overline Z(1), \overline W(1))$, where $\overline X(1)=(X(0), X(1))$ are common causes of subsequent treatment and outcome variables (type 1)), $A(1)$ and $Y$; $\overline Z(1)=(Z(0), Z(1))$ are referred to as sequential treatment-inducing proxies (type 2)); and $\overline W(1)=(W(0),W(1))$ are referred to as sequential outcome-inducing proxies (type 3)) \citep{tchetgen2020introduction}, which are formally defined below.  

We now introduce the class of marginal structural models (MSMs) we wish to make inferences about. Robins and colleagues proposed MSMs \citep{robins1997marginal, robins1999association, robins2000marginal, robins2000marginalb} that encode the joint causal effects of time-varying treatment subject to time-varying confounding. MSMs model the marginal distribution of counterfactual outcomes, possibly conditional on baseline covariates. A marginal structural mean model (MSMM) is an MSM that places restrictions solely on the mean of $Y_{\overline a(1)}$ possibly conditional on baseline variables $V \subset X(0)$, or more formally,
\begin{equation}\label{eq:msmmtwo}
    \E\{Y_{\overline a(1)}|V\} = g\{\overline a(1), V; \beta\},\text{ for any }\overline a(1) \in \cA,
\end{equation}
for a known function $g(\cdot, \cdot; \cdot)$ and $p$-dimensional parameter $\beta$. The truth is defined as $\beta_*$, which we aim to make inferences about. 
For example, a common MSMM is $g\{\overline a(1), V; \beta\} = \mu( \beta_0 + \beta_1 (a(0) + a(1)) + \beta_2 V)$, where $\mu$ corresponds to the inverse of an appropriate choice of link function, e.g. identity link when $Y$ is continuous, or logit link function for a binary outcome. A saturated MSMM corresponds to an MSMM indexed by a parameter of dimension equal to the total number of possible potential outcomes; in the two-occasion setting, a saturated MSMM for instance is given by $\E(Y_{\overline a(1)}) = \beta_0 + \beta_1a(0) + \beta_2a(1) + \beta_3a(0)a(1)$; note that a saturated MSMM is technically a nonparametric model. 
\subsection{Assumptions}
In order to describe our identifying assumptions, suppose for a moment that it is possible to conceptualize joint interventions on $(\overline A(1), \overline Z(1))$, such that the following potential outcomes are well defined  $(Y_{\overline a(1), \overline z(1)}, W_{\overline a(1), \overline z(1)}(1), W_{a(0), z(0)}(0))$ and denote potential outcomes under a hypothetical intervention that sets $\overline A(1)$ and $\overline Z(1)$ to $\overline a(1)$, $\overline z(1)$, respectively. Also, let $\overline U(1) = (U(0), U(1))$ denote time-varying unmeasured variables that confound the causal effect of treatment assigned over time on the outcome measured at the end of follow-up. Throughout, we rely on identification conditions introduced in \cite{tchetgen2020introduction} which we now describe.
\begin{assump}[Sequential Potential Outcome-inducing Confounding Proxies]\label{assump:outproxiestwo}
\begin{equation}\label{eq:wexclutwo}
    W_{\overline a(1), \overline z(1)}(1) = W_{a(0)}(1),~~\forall~\overline a(1), \overline z(1),\text{ almost surely}.
\end{equation}
\begin{equation}
    W_{a(0), z(0)}(0) = W(0),~~\forall~a(0), z(0),\text{ almost surely}.
\end{equation}
\end{assump}
This assumption states that $(A(1), \overline Z(1))$ and $Z(0)$ have no direct effect on $W(1)$ and $W(0)$, respectively. In the MTX study, as suggested in the introduction, because the average number of tender joints at baseline and sixth-month follow-up provide an error prone measurement of underlying disease progression which fully mediates the causal effect of past treatment on the outcome (average number of tender joints at end of follow-up), they may be taken as outcome-inducing confounding proxies. Likewise, a health assessment questionnaire, as a measurement of a patient's evolving health status, prone to recall bias and other forms of measurement error, which may reflect health-seeking behavior as a determinant of treatment initiation, is a good candidate treatment-inducing confounding proxy. 
\begin{assump}[Sequential Potential Treatment-inducing Confounding Proxies]\label{assump:trtproxiestwo}
\begin{equation}\label{eq:zexclutwo}
    Y_{\overline a(1), \overline z(1)} = Y_{\overline a(1)},~~\forall~\overline a(1), \overline z(1)\text{ almost surely}.
\end{equation}
\end{assump}
This implies that $\overline Z(1)$ does not have a direct effect on $Y$ other than through $\overline A(1)$. For example, self-reported health status as measured by a health assessment questionnaire does not by itself cause tender joints, but may determine MTX initiation which in turn may reduce disease progression and subsequent tender joints at end-of follow-up. Furthermore, self-reported health may be associated with disease progression to the extent that it is associated with a patient's underlying health status (e.g. co-morbidities). 

Throughout we make the following standard assumptions: (i) consistency: $Y = Y_{\overline A(1), \overline Z(1)}$, $W(1) = W_{\overline A(1), \overline Z(1)}(1)$, $W(0) = W_{A(0), Z(0)}(0)$ almost surely. That is, a person’s observed outcomes match his/her potential outcomes for the treatment regime he/she did indeed followed; (ii) positivity: $\P(A(1) = a|A(0), \overline U(1), \overline L(1)) > 0$ and $\P(A(0) = a|U(0), L(0)) > 0$ for $a = 1, 0$ almost surely, that is, for any realized history of treatment and covariates (both observed and unobserved) at each follow up time, there is a non-negligible opportunity to receive either treatment. 

\begin{assump}[Sequential Proximal Latent Randomization Assumption]\label{assump:lseqigtwo}
\begin{equation}\label{eq:proxylsitwo}
    \{\overline Z(1), A(1)\} \perp \{W_{a(0), z(0)}(0), W_{\overline a(1), \overline z(1)}(1), Y_{\overline a(1), \overline z(1)}\}~\big|~A(0) = a(0), \overline X(1), \overline U(1),
\end{equation}
\begin{equation}
    \{Z(0), A(0)\} \perp \{W_{a(0), z(0)}(0), Y_{\overline a(1), \overline z(1)}\}~\big|~X(0), U(0).
\end{equation}
\end{assump}
This assumption formally states sequential randomization and thus identifiability of the joint effects of $(\overline Z(1), A(1))$ on $Y$ and $\overline W(1)$ given observed treatment history $A(0)$, covariate history $\overline X(1)$ and unmeasured factors $\overline U(1)$.

Assumptions \ref{assump:outproxiestwo}--\ref{assump:lseqigtwo} together formally define $\overline{Z}(1)$ and $\overline{W}(1)$ as sequential treatment-inducing and outcome-inducing proxies respectively. Technically, the following independence statements implied by Assumptions \ref{assump:outproxiestwo}--\ref{assump:lseqigtwo} can be taken as primitive conditions for our framework, in place of the above assumptions particularly in settings where one does not wish to entertain potential interventions on $\overline Z(1)$.
\begin{equation}\label{eq:proxindZtwo}
    \overline Z(1) \perp Y | \overline A(1), \overline X(1), \overline U(1),
\end{equation}
\begin{equation}\label{eq:proxindWtwo}
    \{\overline Z(1), A(1)\} \perp \overline W(1)~\big|~A(0), \overline X(1), \overline U(1),
\end{equation}
\begin{equation}\label{eq:proxindWtwo2}
    \{Z(0), A(0)\} \perp W(0)~\big|~X(0), U(0).
\end{equation}

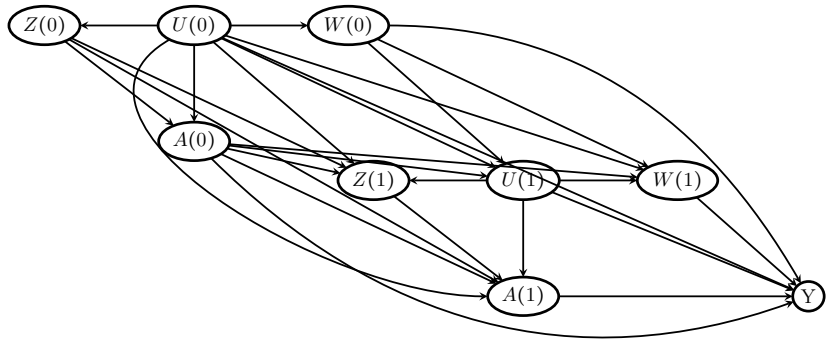
\begin{figure}[H]
    \centering
    \resizebox{.75\textwidth}{!}{


\begin{tikzpicture}
\tikzset{line width=1.5pt, outer sep=0pt,
ell/.style={draw,fill=white, inner sep=2pt,
line width=1.5pt},
swig vsplit={gap=5pt, line color right=red,
inner line width right=0.5pt}};

\node[name=Z0, ell, shape=ellipse] {$Z(0)$};

\node[name=U0, right=15mm of Z0, ell, shape=ellipse]{$U(0)$};

\node[name=W0, right=15mm of U0, ell, shape=ellipse]{$W(0)$};

\node[name=A0, below=15mm of U0, ell, shape=ellipse]{$A(0)$};


\node[name=Z1, below right =35mm of U0, ell, shape=ellipse] {$Z(1)$};

\node[name=U1, right=15mm of Z1, ell, shape=ellipse]{$U(1)$};

\node[name=W1, right=15mm of U1, ell, shape=ellipse]{$W(1)$};

\node[name=A1, below=15mm of U1, ell, shape=ellipse]{$A(1)$};

\node[name=Y, right=45mm of A1, ell, shape=ellipse]{Y};

\draw[->,line width=1.0pt,>=stealth](U0) to (Z0);
\draw[->,line width=1.0pt,>=stealth](Z0) to (A0);
\draw[->,line width=1.0pt,>=stealth](W0) to (U1);

\draw[->,line width=1.0pt,>=stealth](U0) to (W0);
\draw[->,line width=1.0pt,>=stealth](U0) to (A0);
\draw[->,line width=1.0pt,>=stealth](A0) to (W1);

\draw[->,line width=1.0pt,>=stealth](Z0) to (A1);
\draw[->,line width=1.0pt,>=stealth](Z0) to (Z1);

\draw[->,line width=1.0pt,>=stealth](U0) to (Y);

\draw[->,line width=1.0pt,>=stealth](U0) to (Z1);
\draw[->,line width=1.0pt,>=stealth](U0) to [out=210, in=180](A1);
\draw[->,line width=1.0pt,>=stealth](U0) to (W1);
\draw[->,line width=1.0pt,>=stealth](U0) to (U1);
\draw[->,line width=1.0pt,>=stealth](A0) to (A1);
\draw[->,line width=1.0pt,>=stealth](A0) to (U1);
\draw[->,line width=1.0pt,>=stealth](A0) to (Z1);
\draw[->,line width=1.0pt,>=stealth](W0) to (W1);

\draw[->,line width=1.0pt,>=stealth](W0) to [out=0, in=120](Y);

\draw[->,line width=1.0pt,>=stealth](A0) to [out=315, in=200](Y);

\draw[->,line width=1.0pt,>=stealth](U1) to (Z1);
\draw[->,line width=1.0pt,>=stealth](Z1) to (A1);
\draw[->,line width=1.0pt,>=stealth](U1) to (W1);
\draw[->,line width=1.0pt,>=stealth](U1) to (A1);
\draw[->,line width=1.0pt,>=stealth](U1) to (Y);
\draw[->,line width=1.0pt,>=stealth](W1) to (Y);

\draw[->,line width=1.0pt,>=stealth](A1) to (Y);



\end{tikzpicture}

    }
    \caption{A directed acyclic graph with time varying endogenous treatments and time varying proxies when proximal independence assumptions \eqref{eq:proxindZtwo}, \eqref{eq:proxindWtwo} and \eqref{eq:proxindWtwo2} hold. }
    \label{fig:timedep}
\end{figure}

Figure \ref{fig:timedep} illustrates a possible data generating mechanism in which assumptions \ref{assump:outproxiestwo}--\ref{assump:lseqigtwo} and thus \eqref{eq:proxindZtwo}--\eqref{eq:proxindWtwo2} hold, where to simplify the figure time-varying covariates $\overline X(1)$ which are structurally similar to $\overline U(1)$ are suppressed. We also note that alternative DAGs compatible with conditions \eqref{eq:proxindZtwo}, \eqref{eq:proxindWtwo} and \eqref{eq:proxindWtwo2} can in principle be drawn, although throughout, we take Figure \ref{fig:timedep} as a canonical graphical representation of key conditional independence conditions.

\section{Proximal Causal Identification}\label{sec:identwo}
In this section, we describe two approaches for nonparametric proximal identification of the counterfactual mean $\E(Y_{\overline a}|V)$ that will later motivate a rich class of estimating equations for the parameters of an MSMM. It is important to note that the results described below do not presume a particular functional form relating a counterfactual outcome mean to its corresponding treatment regime. We describe two identification results in the time-varying treatment setting, the proximal analog of Robins' g-formula \citep{robins1986new, robins1987graphical, hernan2020causal} obtained by \cite{tchetgen2020introduction} and a novel proximal analog of inverse probability weighting \citep{robins1997marginal, hernan2001marginal, hernan2020causal}. 

\subsection{Identification via Outcome Confounding Bridge Functions}
We first briefly describe the identification result due to \cite{tchetgen2020introduction} based on outcome confounding bridge functions defined as a solution to certain Fredholm integral equations of the first kind. This approach effectively generalizes results due to \cite{miao2018identifying, cui2020semiparametric} for the average treatment effect in the point exposure case to the longitudinal treatment setting. 

The result relies on the following additional conditions codifying an informational relevance requirement the proxies must fulfill.
\begin{assump}[Sequential Proxy Relevance for Outcome Confounding Bridge Functions]\label{assump:outuntestcompletetwo}
~~~
\begin{enumerate}
    \item For any $\overline a(1), \overline x(1)$, and any square-integrable function $\nu$,
\begin{equation}\label{eq:outcompleteuntesttwo}
    \E[\nu\{\overline U(1)\}|\overline a(1), \overline Z(1), \overline x(1)] = 0~\text{if and only if}~\nu\{\overline U(1)\} = 0~\text{almost surely},
\end{equation}
\begin{equation}\label{eq:outcompleteuntesttwo2}
    \E[\nu\{U(0)\}|a(0), Z(0), x(0)] = 0~\text{if and only if}~\nu\{U(0)\} = 0~\text{almost surely}.
\end{equation}
    \item For any $\overline a(1), \overline x(1)$, and any square-integrable function $\nu$,
\begin{equation}\label{eq:treatcompletetesttwo}
    \E[\nu\{\overline Z(1)\}|\overline a(1), \overline W(1), \overline x(1)] = 0~\text{if and only if}~\nu\{\overline Z(1)\} = 0~\text{almost surely},
\end{equation}
\begin{equation}
    \E[\nu\{Z(0)\}|a(0), W(0), x(0)]~\text{if and only if}~\nu\{Z(0)\} = 0~\text{almost surely}.
\end{equation}
\end{enumerate}
\end{assump}

These conditions are formally known as completeness conditions which can accommodate both categorical, discrete and continuous variables. Completeness is essential to ensure existence of a solution to a certain integral equation we consider below, as well as identification of the MSMM. Here one may interpret the first completeness condition \eqref{eq:outcompleteuntesttwo} as a requirement relating the range of $\overline U(1)$ to that of $\overline Z(1)$ which essentially states that the set of proxies must have sufficient variability relative to variability of $\overline U(1)$.
In order to gain intuition about the condition, consider the special case of categorical $\{ U(t), Z(t), W(t):t\}$, with constant cardinality over time $d_u$, $d_z$ and $d_w$ respectively, where the cardinality is defined as the product of the cardinalities of each component in the vector. In this case, completeness requires that
\begin{equation}\label{eq:completecategoricaltwo}
    \min(d_z, d_w) \geq d_u,
\end{equation}
which states that $\overline Z(1)$ and $\overline W(1)$ must each have at least as many categories as $\overline U(1)$. Intuitively, condition \eqref{eq:completecategoricaltwo} states that proximal causal inference can potentially account for unmeasured confounding in the categorical case as long as the number of categories of $\overline U(1)$ is no larger than that of either proxies $\overline Z(1)$ and $\overline W(1)$ \citep{miao2018identifying, shi2020multiply, tchetgen2020introduction, cui2020semiparametric}. Completeness is a familiar technical condition central to the study of sufficiency in the foundational theory of statistical inference. Many commonly-used parametric and semiparametric models such as the semiparametric exponential family \citep{newey2003instrumental} and semiparametric location-scale family \citep{hu2018nonparametric} satisfy the completeness condition. For nonparametric regression models, results of \cite{d2011completeness} and \cite{darolles2011nonparametric} can be used to justify the completeness condition, although their primary focus is on a nonparametric instrumental variable model, where completeness plays a central role. In order to supplement the more succinct discussion given here, a more extensive discussion of the completeness condition is provided in the supplementary material for the interested reader. Also, see \cite{chen2014local}, \cite{andrews2017examples} and references therein for an excellent overview of the role of completeness in nonparametric causal inference. 

\begin{lem}\label{lem:outconftwo}
Under Assumption \ref{assump:outuntestcompletetwo}(b) and regularity Conditions B.1(a, b, c) given in the supplementary material, there exist functions $H_1\{\overline a(1)\} = h_1\{\overline W(1), \overline a(1), \overline X(1)\}$ and $H_0\{\overline a(1)\} = h_0\{W(0), \overline a(1), X(0)\}$ such that 
\begin{equation}\label{eq:outconfbridgeidentwo}
    \E\{Y|\overline a(1), \overline z(1), \overline x(1)\} = \E[H_1\{\overline a(1)\}|\overline a(1), \overline z(1), \overline x(1)],
\end{equation}
and
\begin{equation}\label{eq:outconfbridgeiden2two}
    \E[H_1\{\overline a(1)\}|a(0), z(0), x(0)] = \E[H_0\{\overline a(1)\}|a(0), z(0), x(0)].
\end{equation}
\end{lem}

Equations \eqref{eq:outconfbridgeidentwo} and \eqref{eq:outconfbridgeiden2two} define Fredholm integral equations of the first kind. Lemma \ref{lem:outconftwo} provides sufficient conditions for existence of a solution to these integral equations, however they do not ensure uniqueness of such solutions. Interestingly as noted by \cite{tchetgen2020introduction}, any set of functions $(h_1, h_0)$ satisfying \eqref{eq:outconfbridgeidentwo} and \eqref{eq:outconfbridgeiden2two} uniquely identify $\E(Y_{\overline a(1)}|V)$ as formally stated in the theorem below. A remarkable result of proximal causal inference is that it offers a genuine opportunity to account for $\overline U(k)$ without either measuring $\overline U(k)$ directly or estimating its distribution provided that the set of proxies, though imperfect, is sufficiently rich so that the integral equations \eqref{eq:outconfbridgeidentwo} and \eqref{eq:outconfbridgeiden2two} admit a solution.

\begin{thm}[\cite{tchetgen2020introduction}]\label{thm:outconftwo}
Under assumptions \ref{assump:outproxiestwo}, \ref{assump:trtproxiestwo}, \ref{assump:lseqigtwo} and \ref{assump:outuntestcompletetwo}(a), for $h_1$ and $h_0$ satisfying \eqref{eq:outconfbridgeidentwo} and \eqref{eq:outconfbridgeiden2two}, we have that
\begin{equation}\label{eq:yglutwo}
    \E\{Y_{\overline a(1)}|\overline a(1), \overline u(1), \overline x(1)\} = \E[H_1\{\overline a(1)\}|\overline a(1), \overline u(1), \overline x(1)],
\end{equation}
and
\begin{equation}\label{eq:ygluhktwo}
    \E\{Y_{\overline a(1)}|a(0), u(0), x(0)\} = \E[H_0\{\overline a(1)\}|a(0), u(0), x(0)].
\end{equation}
It follows that 
\begin{equation}\label{eq:causaloutconfbridgegformconditionaltwo}
    \E\{Y_{\overline a(1)}|a(0), \overline x(1)\} = \E[H_1\{\overline a(1)\}|a(0), \overline x(1)],
\end{equation}
\begin{equation}\label{eq:causaloutconfbridgegformconditionaltwotwo}
    \E\{Y_{\overline a(1)}|x(0)\} =\E[H_0\{\overline a(1)\}|x(0)],
\end{equation} and therefore
\begin{equation}\label{eq:causaloutconfbridgegformtwo}
    \E\{Y_{\overline a(1)}|V\} = \E[H_0\{\overline a(1)\}|V] = \E[h_0\{W(0), \overline a(1), X(0)\}|V].
\end{equation}

\end{thm}

\begin{rem}\label{rem:outsratwo}
Under SRA given $\ \overline L(1)=(\overline X(1), \overline W(1))$, such that we may take $\overline Z(1) = \overline W(1)$,
\eqref{eq:outconfbridgeidentwo} and \eqref{eq:outconfbridgeiden2two} simplify to 
\begin{equation}
    h_1\{\overline a(1), \overline l(1)\} = \E\{Y|\overline a(1), \overline l(1)\},
\end{equation}
and
\begin{equation}
    h_0\{\overline a(1), l(0)\} = \E[h_1\{\overline a(1), \overline L(1)\}|a(0), L(0) = l(0)],
\end{equation}
recovering Robins' well-established g-formula \citep{robins1986new, robins1987graphical, hernan2020causal}. 
\end{rem}

\subsection{Identification via Treatment Confounding Bridge Functions}
We now provide new identification results that complement the results given in the last subsection. Specifically, we introduce and leverage so-called treatment confounding bridge functions for identification, an alternative to the outcome confounding bridge function approach. The approach provides a longitudinal generalization of the identification result for the average treatment effect obtained by \cite{cui2020semiparametric}.

Our result relies on an alternative set of completeness conditions.
\begin{assump}[Sequential Proxy Relevance for Treatment Confounding Bridge Functions]\label{assump:treatuntestcompletetwo}
~~~
\begin{enumerate}
    \item For any $\overline a(1), \overline x(1)$, and any square-integrable function $\nu$,
\begin{equation}\label{eq:treatcompleteuntesttwo}
    \E[\nu\{\overline U(1)\}|\overline a(1), \overline W(1), \overline x(1)] = 0~\text{if and only if}~\nu\{\overline U(1)\} = 0~\text{almost surely},
\end{equation}
\begin{equation}
    \E[\nu\{U(0)\}|a(0), W(0), x(0)] = 0~\text{if and only if}~\nu\{U(0)\} = 0~\text{almost surely}.
\end{equation}
    \item For any $\overline a(1), \overline x(1)$, and any square-integrable function $\nu$,
\begin{equation}\label{eq:outcompletetesttwo}
    \E[\nu\{\overline W(1)\}|\overline a(1), \overline Z(1), \overline x(1)] = 0~\text{if and only if}~\nu\{\overline W(1)\} = 0~\text{almost surely},
\end{equation}
\begin{equation}
    \E[\nu\{W(0)\}|a(0), Z(0), x(0)] = 0~\text{if and only if}~\nu\{W(0)\} = 0~\text{almost surely}.
\end{equation}
\end{enumerate}
\end{assump}

\begin{lem}\label{lem:treatconftwo}
Under Assumption \ref{assump:treatuntestcompletetwo}(b) and regularity Conditions B.1(a, d, e) in the supplementary material, there exist functions $Q_0\{a(0)\} = q_0\{Z(0), a(0), X(0)\}$, $Q_1\{\overline a(1)\} = q_1\{\overline Z(1), \overline a(1), \overline X(1)\}$ such that
\begin{equation}\label{eq:treatconfbridgeidentwo}
    \frac{1}{\P\{A(0) = a(0)|w(0), x(0)\}} = \E[Q_0\{a(0)\}|a(0), w(0), x(0)],
\end{equation}
and
\begin{equation}\label{eq:treatconfbridgeiden2two}
    \frac{\E[Q_0\{a(0)\}|a(0), \overline w(1), \overline x(1)]}{\P\{A(0) = a(1)|a(0), \overline w(1), \overline x(1)\}} = \E[Q_1\{\overline a(1)\}|\overline a(1), \overline w(1), \overline x(1)].
\end{equation}
\end{lem}
We then have the following identification result.
\begin{thm}\label{thm:treatconftwo}
Under assumptions \ref{assump:outproxiestwo}, \ref{assump:trtproxiestwo}, \ref{assump:lseqigtwo} and \ref{assump:treatuntestcompletetwo}(a), any functions $q_0$ and $q_1$ satisfying \eqref{eq:treatconfbridgeidentwo} and \eqref{eq:treatconfbridgeiden2two} satisfy
\begin{equation}\label{eq:yqlutwo}
    \frac{1}{\P\{A(0) = a(0)|u(0), x(0)\}} = \E[Q_0\{a(0)\}|a(0), u(0), x(0)],
\end{equation}
and
\begin{equation}\label{eq:yqluqktwo}
    \frac{\E[Q_0\{a(0)\}|a(0), \overline u(1), \overline x(1)]}{\P\{A(1) = a(1)|a(0), \overline u(1), \overline x(1)\}} = \E[Q_1\{\overline a(1)\}|\overline a(1), \overline u(1), \overline x(1)].
\end{equation}
Therefore
\begin{equation}\label{eq:causaltreatconfbridgegformtwo}
    \E\{Y_{\overline a(1)}|V\} = \E[Y\mathbbm{1}\{\overline A(1) = \overline a(1)\}Q_1\{\overline a(1)\}|V] = \E[Y\mathbbm{1}\{\overline A(1) = \overline a(1)\}q_1\{\overline Z(1), \overline a(1), \overline X(1)\}|V].
\end{equation}
\end{thm}

\begin{rem}
Under the SRA considered in Remark \ref{rem:outsratwo}, \eqref{eq:treatconfbridgeidentwo} and \eqref{eq:treatconfbridgeiden2two} simplify to 
\begin{equation}
    q_0\{a(0), l(0)\} = \frac{1}{\P\{A(0) = a(0)|l(0)\}},
\end{equation}
and
\begin{equation}
    q_1\{\overline a(1), \overline l(1)\} = \frac{1}{\P\{a(1)|a(0), \overline l(1)\}\P\{A(0) = a(0)|l(0)\}}.
\end{equation}
recovering standard inverse probability weighting \citep{robins1997marginal}.
\end{rem}

\section{Semiparametric Theory under MSMM}\label{sec:semitwo}

In the previous section, we established that the joint effects of a time-varying treatment can in fact be identified nonparametrically despite unmeasured time-varying confounding, provided proxies satisfy certain conditions. In principle one may wish to estimate the treatment effects under a nonparametric MSMM, however, in practice in order to manage the curse of dimensionality, it is customary to conduct inferences under a parametric or semiparametric MSMM. In this section, we derive the set of regular and asymptotically linear estimators and the efficiency lower bound of the parameters of an MSMM under a semiparametric model which is otherwise unrestricted. Note that the MSMM restriction \eqref{eq:msmmtwo}, is equivalent to the moment restriction that for any $p$-dimensional measurable functions $d\{\overline A(1), V\}$,
\begin{equation}
    \E\left(\sum_{\overline a(1) \in \cA}d\{\overline a(1), V\}[\E(Y_{\overline a(1)}|V) - g\{\overline a(1), V; \beta\}]\right) = 0.
\end{equation}
Clearly, these moment equations cannot be evaluated empirically and are therefore infeasible due to dependence on potential outcomes that are not observable, however, under Theorem \ref{thm:outconftwo} an observable analog of these moment equations can be obtained, mainly:
\begin{equation}\label{eq:observedmsmmtwo}
    \E\{D(\beta, d)\} = 0,
\end{equation}
where $D(\beta, d) = \sum_{\overline a(1) \in \cA}d\{\overline a(1), V\}[H_0\{\overline a(1)\} - g\{\overline a(1), V; \beta\}]$ and $H_0\{\overline a(1)\}$ is the outcome confounding bridge function defined in \eqref{eq:outconfbridgeiden2two}.
To proceed with inference, let ${\cal M}$ denote the semiparametric model consisting of all observed data distributions for which integral equations \eqref{eq:outconfbridgeidentwo}, \eqref{eq:outconfbridgeiden2two} admit a solution that satisfies the MSMM for the proximal g-formula given by \eqref{eq:observedmsmmtwo}. Note that this semiparametric model is quite rich, including data-generating mechanisms for which the conditions of Lemma \ref{lem:outconftwo} hold. Let $L^1(S)$ and $L^2(S)$ be spaces of all integrable and all square-integrable functions of a random variable $S$, respectively. That is,
\begin{equation}
    L^1(S) = \left\{f: \int |f(S)| d\P(S) < \infty\right\},
\end{equation}
\begin{equation}
    L^2(S) = \left\{f: \int f^2(S) d\P(S) < \infty\right\}.
\end{equation}
Define $T_1: L^2\{\overline W(1), A(0), \overline X(1)\} \to L^2\{Z(0), A(0), X(0)\}$, $T_0: L^2\{W(0), A(0), X(0)\} \to L^2\{Z(0), A(0), X(0)\}$ as the $L^2$ extension of conditional expectation operators, namely, when restricting $T_1$ to $s \in L^1\{\overline W(1), A(0), \overline X(1)\} \cap L^2\{\overline W(1), A(0), \overline X(1)\}$,
\begin{equation}
    T_1(s) = \E[s\{\overline W(1), A(0), \overline X(1)\}|Z(0), A(0), X(0)],
\end{equation}
when restricting $T_0$ to $s \in L^1\{W(0), A(0), X(0)\} \cap L^2\{W(0), A(0), X(0)\}$,
\begin{equation}
    T_0(s) = \E[s\{W(0), A(0), X(0)\}|Z(0), A(0), X(0)].
\end{equation}
\begin{equation}
    \cT_1 = \{S \in L_{2, 0}(\cO): \E\{(Y - H_1)S(\cO)|\overline a, \overline z, \overline x\} \in \text{range}(T_1)\},
\end{equation}
and
\begin{equation}
    \cT_0 = \{S \in L_{2, 0}(\cO): \E\{(H_1 - H_0)S(\cO)|a(0), z(0), x(0)\} \in \text{range}(T_0)\}.
\end{equation}

\begin{thm}\label{thm:raltwo}
Any regular and asymptotically linear estimator $\widehat \beta$ of $\beta_*$ in $\cal M$, at a law where \eqref{eq:treatconfbridgeidentwo} and \eqref{eq:treatconfbridgeiden2two} admit a solution, must satisfy the following:
\begin{align}
   n^{1/2}(\widehat \beta - \beta_*)= n^{1/2}\P_n(S) + o_p(1),
\end{align}
where
\begin{equation}\label{eq:generalclosuretwo}
    S \in \text{closure}\left[\left\{\{k(d)\}^{-1}R(\beta_*, d) + \cT_1^\perp + \cT_0^\perp: \text{ for any } d\right\}\right].
\end{equation}
\end{thm}

The theorem provides a characterization of all influence functions of regular and asymptotically linear estimators of MSMM parameters. An important special case we later leverage, arises when the following assumption holds.
\begin{assump}\label{assump:ralregularitytwo}
$T_1$ and $T_0$ are surjective.
\end{assump}
The assumption essentially states that $L^2\{\overline W(1), A(0), \overline X(1)\}$ and $L^2\{W(0), A(0), X(0)\}$ are sufficiently rich so that mapping them onto $L^2\{Z(0), A(0), X(0)\}$ via conditional expectation operator can generate all elements of the latter space.
\begin{cor}\label{cor:raltwo}
Any regular and asymptotically linear estimator $\widehat \beta$ of $\beta_*$ in $\cal M$, at a law where \eqref{eq:treatconfbridgeidentwo} and \eqref{eq:treatconfbridgeiden2two} admit a solution and Assumption \ref{assump:ralregularitytwo} holds, must satisfy the following:
\begin{align}
   n^{1/2}( \widehat \beta- \beta_*)= n^{1/2}\{k(d)\}^{-1}\P_n\{R(\beta_*, d)\} +o_p(1),
\end{align} 
where $\P_n$ is the sample mean, 
\begin{equation}
    R(\beta, d) = \sum_{\overline a(1) \in \cA}d\{\overline a(1), V\}\Xi(\beta)_{\overline a(1)},
\end{equation}
for some $p$-dimensional measurable function $d\{\overline A(1), V\}$, 
\begin{align}
    \Xi(\beta)_{\overline a(1)} &:= \mathbbm{1}\{\overline A(1) = \overline a(1)\}Q_1\{\overline a(1)\}[Y - H_1\{\overline a(1)\}]\\
    &~~~+\mathbbm{1}\{A(0) = a(0)\}Q_{0}\{a(0)\}[H_1\{\overline a(1)\} - H_0\{\overline a(1)\}]\\
    &~~~ + H_0\{\overline a(1)\} - g\{\overline a(1), V; \beta\}],
\end{align} 
and
\begin{equation}
    k(d) = -\E\left\{\frac{\partial R(\beta_*, d)}{\partial \beta}\right\} = -\E\left\{\frac{\partial D(\beta_*, d)}{\partial \beta}\right\}.
\end{equation} 
Furthermore, the optimal index $d_{\text{eff}}$ of $d$ and thus the semiparametric efficiency bound for ${\cal M}$ are given by equations (B.15) and (B.16) of the supplementary material.
\end{cor}

\section{Proximal Estimation under MSMM}\label{sec:esttwo}
It is straightforward to prove that $\widehat{\beta}$ in Theorem \ref{thm:raltwo} can in fact be obtained (up to asymptotic equivalence) under the conditions given in the theorem by solving $\P_n[R(\beta, d)] = 0$; however, clearly such an estimator is technically not feasible as it depends crucially on complicated functions of the true (unknown) observed data distribution, mainly $(h_1, h_0)$ and $(q_1, q_0)$ that satisfy \eqref{eq:outconfbridgeidentwo}, \eqref{eq:outconfbridgeiden2two} and \eqref{eq:treatconfbridgeidentwo}, \eqref{eq:treatconfbridgeiden2two}. Empirical solutions to these integral equations are notoriously challenging to compute due to the ill-posedness nature of the problem, typically requiring a form of regularization. In the point treatment case, parametric \citep{tchetgen2020introduction}, semiparametric \citep{shi2020multiply, miao2018confounding, cui2020semiparametric}, and nonparametric approaches \citep{cui2020semiparametric, shi2020multiply, kallus2021causal, ghassami2022minimax, mastouri2021proximal, deaner2018panel} have recently been considered for estimation and inference about causal effects using the proximal framework. The results of \cite{ghassami2022minimax} suggest that root-$n$ estimation of $\beta_*$ may not be attainable even in the point treatment case in moderate to high dimensional settings primarily due to necessarily slow convergence rates of nonparametric estimation of confounding bridge functions, further aggravated by the potential ill-posedness of moment equations defining them. Our current longitudinal setting is considerably more challenging than considered in these prior works, as the number of bridge functions and their dimensionality expand significantly over time, rendering nonparametric estimation practically infeasible. In order to resolve this difficulty, we consider a practical approach to constructing feasible moment estimating equations for $\beta_*$ under low dimensional smooth working models for the nuisance parameters $(h_1, h_0, q_1, q_0)$. Nevertheless, as we establish below we can mitigate concerns about model dependence to some extent, as our approach enjoys some degree of robustness against misspecification of working models for confounding bridge functions as our moment equations for the underlying MSMM of primary scientific interest are in fact doubly robust.

We first detail our proposed approach to estimate $(h_1, h_0, q_1, q_0)$. Let $h_1(\cdot) = h_1(\cdot; b_1)$, $h_0(\cdot) = h_0(\cdot; b_0)$, $q_1(\cdot) = q_1(\cdot; t_1)$, $q_0(\cdot) = q_0(\cdot; t_0)$ denote parametric working models indexed by low dimensional parameters $(b_1, b_0, t_1, t_0)$, respectively. Let $M_1 = m_1(\overline Z(1)$, $\overline A(1)$, $\overline X(1))$, $M_0 = m_0(Z(0)$, $A(0)$, $X(0))$ for some measurable functions $m_1()$, $m_0()$ that are of the same dimensions as $b_1$ and $b_0$. Also, write $N_1 = n_1(\overline W(1), \overline A(1), \overline X(1))$, $N_0 = n_0(W(0)$, $A(0)$, $X(0))$ for some measurable function $n_1()$, $n_0()$ that are of the same dimensions as $t_1$ and $t_0$. We also denote $N_{1, +} = n_1(\overline W(1), A(1) = 1, A(0), \overline X(1)) + n_1(\overline W(1), A(1) = 0, A(0), \overline X(1))$ and $N_{0, +} = n_0(W(0), A(0) = 1, X(0)) + n_0(W(0), A(0) = 0, X(0))$. By Theorem E.1 in the supplementary material, estimators of $(\widehat h_1, \widehat h_0, \widehat q_1, \widehat q_0)$ can be obtained by solving the following estimating equations
\begin{equation}\label{eq:h1est}
    \P_n([Y - H_1\{\overline A(1); b_1\}]M_1) = 0,
\end{equation}
\begin{equation}\label{eq:h0est}
    \P_n([H_{1}\{a(1), A(0); \widehat b_{1}\} - H_{0}\{a(1), A(0); b_{0}\}]M_0) = 0,
\end{equation}
\begin{equation}\label{eq:q0est}
    \P_n[Q_{0}\{A(0); t_{0}\}N_0 - N_{0, +}] = 0,
\end{equation}
\begin{equation}\label{eq:q1est}
    \P_n[Q_{1}\{\overline A(1); t_{1}\}N_1 - Q_{0}\{A(0); \widehat t_{0}\}N_{1, +}] = 0.
\end{equation}
Simple working models of parameterization of confounding bridge are as following
\begin{equation}\label{eq:simuh1}
    h_1\{\overline W(1), \overline A(1), \overline X(1); b_1\} = b_{1, 0} + b_{1, a}^\top \overline A(1) + b_{1, w}^\top \overline W(1) + b_{1, x}^\top \overline X(1),
\end{equation}
\begin{equation}\label{eq:simuh0}
    h_0\{W(0), \overline A(1), X(0); b_0\} = b_{0, 0} + b_{0, a}^\top\overline A(1) + b_{0, w} W(0) + b_{0, x}X(0),
\end{equation}
\begin{equation}\label{eq:simuq0}
    q_0\{Z(0), A(0), X(0); t_0\} = 1 + \exp\left[(-1)^{1 - A(0)}\{t_{0, 0} + t_{0, a}A(0) + t_{0, z} Z(0) + t_{0, x}X(0)\}\right],
\end{equation}
and
\begin{align}
    &q_1\{\overline Z(1), \overline A(1), \overline X(1); t_1\} \\
    &= 1 + q_0\{Z(0), A(0), X(0); t_0\} \\
    &+ \exp[(-1)^{1 - A(1)}\{t_{1, 0} + t_{1, a}^\top \overline A(1) + t_{1, z}\overline Z(1) + t_{1, x}^\top \overline X(1)\}]\\
    & + q_0\{Z(0), A(0), X(0); t_0\}\exp[(-1)^{1 - A(1)}\{t_{1, 0} + t_{1, a}^\top \overline A(1) + t_{1, z}\overline Z(1) + t_{1, x}^\top \overline X(1)\}].  \label{eq:simuq1}
\end{align}
Under SRA considered in Remark \ref{rem:outsratwo}, the working models considered above correspond to imposing linear models on $\E\{Y|\bar A(1), \bar L(1)\}$, $\E[\E\{Y|a(1), A(0), \bar L(1)\}|A(0), L(0)]$, and logistic models on $\P\{A(1)|A(0), \bar L(1)\}$, $\P\{A(0)|L(0)\}$. Our selection of working models generalizes those considered in \citet{cui2020semiparametric}. In this case, a natural choice sets $M_1 = \{1, \overline A(1), \overline Z(1), \overline X(1)\}^\top$, $M_0 = \{1, A(0), Z(0), X(0)\}^\top$ and $N_1 = (-1)^{1 - A(1)}\{1, \overline A(1), \overline W(1), \overline X(1)\}^\top$, $N_0= (-1)^{1 - A(0)}\{1, A(0), W(0), X(0)\}^\top$. In fact, \eqref{eq:h1est}, \eqref{eq:h0est}, \eqref{eq:q0est} and \eqref{eq:q1est} become
\begin{equation}
    \P_n[\{Y - H_1(\overline A(1); b_1)\}\{1, \overline Z(1), \overline A(1), \overline X(1)\}^\top] = 0,
\end{equation}
\begin{equation}
    \P_n([H_{1}\{a(1), A(0); \widehat b_{1}\} - H_{0}\{a(1), A(0); b_{0}\}]\{1, Z(0), A(0), X(0)\}^\top) = 0.
\end{equation}
\begin{equation}
    \P_n[(-1)^{1 - A(0)}Q_{0}\{A(0); t_{0}\}\{1, W(0), A(0), X(0)\}^\top - \{0, (0)_{p_{w(0)}}, 1, (0)_{p_{x(0)}}\}^\top] = 0,
\end{equation}
\begin{align}
    &\P_n((-1)^{1 - A(1)}Q_{1}\{\overline A(1); t_{1}\}\{1, \overline W(1), A(0), A(1), \overline X(1)\}^\top \\
    &~~~- [0, (0)_{p_{\overline w(1)}}, 0, Q_{0}\{A(0); \widehat t_{0}\}, (0)_{p_{\overline x(1)}}]^\top) = 0,
\end{align}
where $p_{w(0)}$, $p_{x(0)}$, $p_{\overline w(1)}$ and $p_{\overline x(1)}$ denote dimension of $W(0)$, $X(0)$, $\overline W(1)$ and $\overline X(1)$, respectively.

Resulting estimators $(\widehat h_1, \widehat h_0, \widehat q_1, \widehat q_0)$ can then be used to construct corresponding substitution estimators of $\beta_*$. We describe three practical classes of estimators for estimating $\beta_*$. Specifically, the first approach entails a large class of proximal outcome regression estimators (POR) $\widehat \beta_{\text{POR}} = \widehat \beta_{\text{POR}}(d)$ of $\beta_*$ defined as solution to
\begin{equation}
    \P_n\left(\sum_{\overline a(1) \in \cA}d\{\overline a(1), V\}[\widehat H_0\{\overline a(1)\} - g\{\overline a(1), V; \beta\}]\right) = 0.\label{eq:porest}
\end{equation}
The second class of estimators entail a large class of proximal inverse probability weighted estimators (PIPW) $\widehat \beta_{\text{PIPW}} = \widehat \beta_{\text{PIPW}}(d)$ of $\beta_*$ defined as solution to
\begin{equation}
    \P_n\left(d\{\overline A(1), V\} \widehat Q_1\{\overline A(1)\}[Y - g\{\overline A(1), V; \beta\}]\right) = 0.\label{eq:pipwest}
\end{equation}

When $\overline L(1)$ is high dimensional, one cannot be confident that either set of working model $(h_1, h_0)$ or $(q_1, q_0)$ can be specified correctly, a prerequisite for consistent estimation of the MSMM. It is therefore of interest to develop doubly robust estimators of MSMMs, under parametric/semiparametric restrictions on confounding bridge functions, which are guaranteed to deliver valid inferences about $\beta_*$ provided that one but not necessarily both low dimensional working models used to estimate $(h_1, h_0)$ and $(q_1, q_0)$ can be specified correctly. To this end, motivated by Theorem \ref{thm:raltwo}, a class of proximal doubly robust estimators (PDR) $\widehat \beta_{\text{PDR}} = \widehat \beta_{\text{PDR}}(d)$ of $\beta_*$ is obtained as solution to estimating equations of form:
\begin{equation}
    \P_n\left[\sum_{\overline a(1) \in \cA}d\{\overline a(1), V\}\widehat \Xi(\beta)_{\overline a(1)}\right] = 0,\label{eq:pdrest}
\end{equation}
where
\begin{align}
    \widehat \Xi(\beta)_{\overline a(1)} &:= \mathbbm{1}\{\overline A(1) = \overline a(1)\}\widehat Q_1\{\overline a(1)\}[Y - \widehat H_1\{\overline a(1)\}]\\
    &~~~+ \mathbbm{1}\{A(0) = a(0)\}\widehat Q_{0}(a(0))[\widehat H_1\{\overline a(1)\} - \widehat H_0\{\overline a(1)\}]\\
    &~~~ + \widehat H_0\{\overline a(1)\} - g\{\overline a(1), V; \beta\}.
\end{align}
An algorithmic summary of the steps towards constructing the above estimators is given in Algorithm \ref{al:esttwo}.

{\small
\begin{algorithm}[H]
\SetAlgoLined
Step 1: Nuisance parameters estimation:\\
Solve
\begin{equation}
    \P_n([Y - H_1\{\overline A(1); b_1\}]\{1, \overline Z(1), \overline A(1), \overline X(1)\}^\top) = 0,
\end{equation}
\begin{equation}
    \P_n[(-1)^{1 - A(0)}Q_0(A(0); t_0)\{1, W(0), A(0), X(0)\}^\top - \{0, (0)_{p_{w(0)}}, 1, (0)_{p_{x(0)}}\}^\top] = 0,
\end{equation}
to get estimates $\widehat b_1$ and $\widehat t_0$. Then solve
\begin{equation}
    \P_n([H_1\{a(1), A(0); \widehat b_1\} - H_0\P\{a(1), A(0); b_0\}]\{1, Z(0), A(0), X(0)\}^\top) = 0.
\end{equation}
\begin{align}
    &\P_n((-1)^{1 - A(1)}Q_1\{\overline A(1); t_1\}\{1, \overline W(1), A(0), A(1), \overline X(1)\}^\top \\
    &~~~- [0, (0)_{p_{\overline w(1)}}, 0, Q_0\{A(0); \widehat t_0\}, (0)_{p_{\overline x(1)}}]^\top) = 0.
\end{align}
to get estimates $\widehat b_0$ and $\widehat t_1$. These yield $(\widehat h_1, \widehat h_0)$ and $(\widehat q_1, \widehat q_0)$.\\
Step 2: Estimation of MSMM parameter:\\
$\widehat \beta_{\text{POR}}(d)$ is a solution to
\begin{equation}
    \P_n\left[\sum_{\overline a(1) \in \cA}d\{\overline a(1), V\}(\widehat H_0\{\overline a(1)\} - g\{\overline a(1), V; \beta\})\right] = 0.
\end{equation}
$\widehat \beta_{\text{PIPW}}(d)$ is a solution to
\begin{equation}
    \P_n[d\{\overline A(1), V\} \widehat Q_1(\overline A(1))(Y - g(\overline A(1), V; \beta))] = 0,
\end{equation}
and $\widehat \beta_{\text{PDR}}(d)$ is a solution to 
\begin{equation}
    \P_n\left[\sum_{\overline a(1) \in \cA}d\{\overline a(1), V\}\widehat \Xi(\beta)_{\overline a(1)}\right] = 0.
\end{equation}
\caption{Computation of the POR, PIPW and PDR estimators}
\label{al:esttwo}
\end{algorithm}
}
\begin{rem}
In step 1 of the Algorithm \ref{al:esttwo}, one might, for instance, find $\widehat b_1$ by minimizing a corresponding square loss function, that is, letting
\begin{equation}
    S_n = \P_n[\{Y - H_1(\overline A(1); b_1)\}\{1, \overline Z(1), \overline A(1), \overline X(1)\}^\top] = 0,
\end{equation}
$\widehat b_1$ can be defined as the minimizer of
\begin{equation}
    \min S_nS_n^\top.
\end{equation}
This can readily be implemented with the ``optim'' function of the ``stats'' package in R. In step 2 of the Algorithm \ref{al:esttwo}, if a linear MSMM is imposed on $g$ as in our simulations and data application, one can simply run (weighted) linear regressions to compute $\hat \beta$ by setting $d = g$.
\end{rem}

The following theorem provides the asymptotic behavior of our proposed estimators, using standard large sample arguments.
Let ${\cal M}_h$ denote the collection of observed data generating laws under which specified working models $(h_1(\cdot; b_1), h_0(\cdot; b_0))$ are correctly specified, and the model is otherwise unrestricted; likewise, let ${\cal M}_q$ denote the collection of observed data laws under which $(q_1(\cdot; t_1), q_0(\cdot; t_0))$ are correctly specified with unknown parameters $(b_1, b_0)$ and $(t_1, t_0)$ respectively. Specifically,
\begin{align*}
    {\cal M}_h&=\{ h_1\{\overline W(1), \overline A(1) \overline X(1)\} = h_1\{\overline W(1), \overline A(1) \overline X(1); b_1\}, \text{ for some value of } b_1, \\
    & h_0\{W(0), \overline A(1), X(0)\} = h_0\{W(0), \overline A(1), X(0); b_0\}, \text{ for some value of } b_0, \\
&\text{such that \eqref{eq:outconfbridgeidentwo} and \eqref{eq:outconfbridgeiden2two} hold}\};
\end{align*}
\begin{align*}
    {\cal M}_q&=\{ q_1\{\overline Z(1), \overline A(1), \overline X(1)\} = q_1\{\overline Z(1), \overline A(1), \overline X(1); t_1\},\text{ for some value of }t_1,\\
    & q_0\{Z(0), A(0), X(0)\} = q_0\{Z(0), A(0), X(0); t_0\}, \text{ for some value of }t_0,\\
&\text{such that \eqref{eq:treatconfbridgeidentwo} and \eqref{eq:treatconfbridgeiden2two} hold}\};
\end{align*}

\begin{thm}\label{thm:doublyrobusttwo}
Under Assumptions \ref{assump:outproxiestwo}--\ref{assump:treatuntestcompletetwo}, the estimators $\widehat \beta_{\text{POR}}$, $\widehat \beta_{\text{PIPW}}$ and $\widehat \beta_{\text{PDR}}$ are consistent for $\beta_*$ and asymptotically normal under MSMM \eqref{eq:msmmtwo} and  ${\cal M}_h$, ${\cal M}_q$ and ${\cal M}_h \cup {\cal M}_q$, respectively.
\end{thm}

We derive the asymptotic distribution of various proposed estimators which may be used to obtain inferences about the MSMM. 

(a) Suppose $n^{1/2}(\hat b_0 - b_0^*) = n^{-1/2}\sum_{i = 1}^n\eps_{b_0^*, i} + o_p(1)$ and $b_0^*$ is the truth, which holds under ${\cal M}_h$, the POR $\hat \beta_{\text{POR}}$ obtained by solving 
\begin{equation}
    n^{1/2}\{D(\beta, d, \hat b_0)\} = n^{1/2}\P_n\left[\sum_{\overline a \in \cA}d(\overline a, V)\{\widehat H_0(\overline a; \hat b_0) - g(\overline a, V; \beta)\}\right] = o_p(1),
\end{equation} 
is consistent and asymptotically normal for $\beta_*$, with influence function $\text{IF}_{POR, i}$ given by
\begin{align}
    &\sqrt{n}(\hat \beta_{\text{POR}} - \beta_*) = \frac{1}{\sqrt{n}}\sum_{i = 1}^n\text{IF}_{POR, i} + o_p(1)\\
    &= \frac{1}{\sqrt{n}}\sum_{i = 1}^n\left[-\P\left\{\frac{\partial D(\beta_*, d, b_0^*)}{\partial \beta}\right\}\right]^{-1}\left[D_i(\beta_*, d, b_0^*) + \P\left\{\frac{\partial D(\beta_*, d, b_0^*)}{\partial b_0}\right\}\eps_{b_0^*, i}\right]+ o_p(1).
\end{align}

(b) Suppose $n^{1/2}(\hat t_K - t_K^*) = n^{-1/2}\sum_{i = 1}^n\eps_{t_K^*, i} + o_p(1)$ and $t_K^*$ is the truth, which holds under ${\cal M}_q$, the PIPW $\hat \beta_{\text{PIPW}}$ obtained by solving 
\begin{equation}
    n^{1/2}\P_n\{D(\beta, d, \hat q_K)\} = n^{1/2}\P_n\left[d(\overline A, V) \widehat Q_K(\overline A)\{Y - g(\overline A, V; \beta)\}\right] = o_p(1),
\end{equation} 
is consistent and asymptotically normal for $\beta_*$, with influence function $\text{IF}_{PIPW, i}$ given by
\begin{align}
    &\sqrt{n}(\hat \beta_{\text{PIPW}} - \beta_*) = \frac{1}{\sqrt{n}}\sum_{i = 1}^n\text{IF}_{PIPW, i} + o_p(1)\\
    &= \frac{1}{\sqrt{n}}\sum_{i = 1}^n\left[-\P\left\{\frac{\partial D(\beta_*, d, q_K^*)}{\partial \beta}\right\}\right]^{-1}\left[D_i(\beta_*, d, q_K^*) + \P\left\{\frac{\partial D(\beta_*, d, q_K^*)}{\partial t_1}\right\}\eps_{t_K^*, i}\right] + o_p(1).
\end{align}

(c) Suppose $n^{1/2}(\hat b_0 - b_0^*) = n^{-1/2}\sum_{i = 1}^n\eps_{b_0^*, i} + o_p(1)$, $n^{1/2}(\hat t_K - t_K^*) = n^{-1/2}\sum_{i = 1}^n\eps_{t_K^*, i} + o_p(1)$ and either $\{b_k^*\}$ or $\{t_k^*\}$ is the truth, the PDR $\hat \beta_{\text{PDR}}$ obtained by solving 
\begin{equation}
    n^{1/2}\P_n\{R(\beta, d, \hat h, \hat q)\} = n^{1/2}\P_n\left\{\sum_{\overline a \in \cA}d(\overline a, V)\widehat \Xi(\beta)_{\overline a}\right\} = o_p(1),
\end{equation}  
is consistent and asymptotically normal for $\beta_*$, with influence function $\text{IF}_{PDR, i}$ given by
\begin{align}
    &\sqrt{n}(\hat \beta_{\text{PDR}} - \beta_*) = \frac{1}{\sqrt{n}}\sum_{i = 1}^n\text{IF}_{PDR, i} + o_p(1) \\
    &= \frac{1}{\sqrt{n}}\sum_{i = 1}^n\left[-\P\left\{\frac{\partial R(\beta_*, d, h^*, q^*)}{\partial \beta}\right\}\right]^{-1}\bigg[R_i(\beta_*, d, h^*, q^*) \\
    &~~+ \P\left\{\frac{\partial R(\beta_*, d, h^*, q^*)}{\partial b_0}\right\}\eps_{b_0^*, i} + \P\left\{\frac{\partial R(\beta_*, d, h^*, q^*)}{\partial b_1}\right\}\eps_{b_1^*, i} \\
    &~~+ \P\left\{\frac{\partial R(\beta_*, d, h^*, q^*)}{\partial t_0}\right\}\eps_{t_0^*, i} + \P\left\{\frac{\partial R(\beta_*, d, h^*, q^*)}{\partial t_1}\right\}\eps_{t_1^*, i}\bigg] + o_p(1).
\end{align}
The empirical variance covariance matrices $\P_n(\text{IF}_{POR}\text{IF}_{POR}^\top)$, $\P_n(\text{IF}_{PIPW}\text{IF}_{PIPW}^\top)$, and $\P_n(\text{IF}_{PDR}\text{IF}_{PDR}^\top)$ can be used to provide variance estimates for $\hat \beta_{\text{POR}}$, $\hat \beta_{\text{PIPW}}$, and $\hat \beta_{\text{PDR}}$ by replacing the true parameters and $\P$ in $\text{IF}_{POR}$, $\text{IF}_{PIPW}$ and $\text{IF}_{PDR}$ by their estimates and $\P_n$.

\section{Causal Effects of Methotrexate on Rheumatoid Arthritis}\label{sec:real}
We reanalyze data from \cite{choi2002methotrexate} on the potential protective effects of the anti-rheumatic therapy Methotrexate (MTX) among patients with rheumatoid arthritis. While \cite{choi2002methotrexate} focused on survival as an endpoint and using a marginal structural Cox model to quantify joint treatment effects under SRA, here we consider the joint causal effects of MTX on the average of the reported number of tender joints under an MSMM, a crucial measure of disease progression, without appealing to SRA. These causal effects were also examined in \cite{tchetgen2020introduction} by employing a proximal recursive least squares algorithm, a proximal g-computation algorithm based on linear outcome confounding bridge functions specification.

A thousand and ten patients with rheumatoid arthritis met our inclusion criteria, 183 of them were treated with MTX after six months of follow-up. We have recorded baseline covariates including age, sex, education level, rheumatoid arthritis duration and rheumatoid factor positive (rapos). Time varying covariates include current smoking status (smoking), health assessment questionnaire (haqc), number of tender joints (jc), patient's global assessment (gsc), erythrocyte sedimentation rate (esrc), number of disease modifying antirheumatic drugs taken (dmrd) and prednisone use (onprd2) at baseline and sixth month. The treatment process of interest is defined as use of MTX at baseline and month-six of follow-up. As in \cite{choi2002methotrexate}, MTX initiation defines exposure status i.e., once a patient starts MTX therapy, he or she was considered on therapy for the rest of the follow-up. This approach provides a conservative estimate of MTX efficacy just as intent-to-treat analysis does in randomized clinical trial. Therefore the possible treatment strategies are $\cA = \{(0, 0), (0, 1), (1, 1)\}$. Similar to \cite{tchetgen2020introduction}, outcome is defined as the average of reported number of tender joints at month-twelve of follow-up. 

We selected proxies from available time-varying covariates; excluding dmrd and onprd2 as both are antirheumatic treatments which are more likely to have direct effects on both MTX initiation and disease progression. Candidates proxies included smoking status, haqc, jc, gsc, esrc. Our allocation of covariates to various bucket types was consistent with that of \cite{tchetgen2020introduction}, mainly:
\begin{itemize}
    \item $\overline X(1)$ = (age, education, sex, smoking, rheumatoid arthritis duration, rheumatoid factor positive (rapos), prednisone use (onprd2), number of disease modifying antirheumatic drugs taken (dmrd)), where smoking, dmrd and onprd2 are time varying;
    \item $\overline Z(1)$ = (health assessment questionnaire (haqc), erythrocyte sedimentation rate (esrc));
    \item $\overline W(1)$ = (number of tender joints (jc), patient's global assessment (gsc)).
\end{itemize}

We specified the MSMM
\begin{equation}\label{eq:saturatedrealtwo}
    \E(Y_{\overline a(1)}) = \beta_0 + \beta_1 (1 - a(0))a(1) + \beta_2 a(0),
\end{equation}
which is a saturated MSMM. By this definition, $\beta_1$ and $\beta_2$ encode the causal effect of MTX starting on the sixth month and baseline, respectively.

We estimated $\beta_1, \beta_2$ by POR \eqref{eq:porest}, PIPW \eqref{eq:pipwest} and PDR \eqref{eq:pdrest} with working models for $h_1$, $h_0$, $q_0$, and $q_1$ specified as \eqref{eq:simuh1}, \eqref{eq:simuh0}, \eqref{eq:simuq0} and \eqref{eq:simuq1}, as in the simulations reported in the supplementary material. Note that as this MSMM is nonparametric, the PDR estimator is fully efficient at the intersection submodel where all confounding bridge functions are correctly specified. In addition to proximal causal inference, for comparison, we estimated $\beta_1, \beta_2$ via the standard doubly robust estimator (DR) assuming SRA conditional on all baseline and time-varying covariates. The results are given in Table \ref{tab:realresults}.

\begin{table}
\caption{\label{tab:realresults}
Results of real data application for saturated MSMM \eqref{eq:saturatedrealtwo} and cumulative effect MSMM \eqref{eq:nonsaturatedreal}. We report point estimates from POR, PIPW, PDR and standard DR, together with their 95\% confidence intervals in parentheses.
}
\centering
\begin{tabular}{|l|l|}
\hline
\multirow{8}{*}{The saturated MSMM \eqref{eq:saturatedrealtwo}} 
&$\widehat \beta_{1, \text{POR}} = -0.21 (-0.41, 0.01)$ \\ 
&$\widehat \beta_{2, \text{POR}} = -0.47 (-0.67, -0.28)$  \\ \cline{2-2}
& $\widehat \beta_{1, \text{PIPW}} = -0.29 (-1.12, 0.54)$    \\ 
&$\widehat \beta_{2, \text{PIPW}} = -0.44 (-0.80, -0.07)$ \\ \cline{2-2}
&$\widehat \beta_{1, \text{PDR}} = -0.34 (-1.11, 0.43)$  \\
&$\widehat \beta_{2, \text{PDR}} = -0.58 (-0.84, -0.31)$ \\ \cline{2-2}
&$\widehat \beta_{1, \text{DR}} = -0.12 (-0.91, 1.41)$   \\ 
&$\widehat \beta_{2, \text{DR}} = -0.34 (-0.91, 0.22)$  \\ \hline
\multirow{4}{*}{The cumulative MSMM \eqref{eq:nonsaturatedreal}} 
&$\widehat \beta_{1, \text{POR}} = -0.24 (-0.33, -0.14)$     \\ \cline{2-2}
&$\widehat \beta_{1, \text{PIPW}} = -0.22 (-0.41, -0.03)$\\ \cline{2-2}
&$\widehat \beta_{1, \text{PDR}} = -0.29 (-0.42, -0.16)$ \\ \cline{2-2}
&$\widehat \beta_{1, \text{DR}} = -0.17 (-0.41, 0.06)$         \\ \hline
\end{tabular}
\end{table}

Point estimates from POR, PIPW and PDR are consistent with each other and therefore by double robustness, there is no evidence of model misspecification. Results reflected by all three proximal estimators indicate a significant protective effect of MTX against disease progression when treatment is initiated at baseline. The corresponding doubly robust SRA-based estimator suggests a substantially smaller protective effect of MTX which fails to meet statistical significance. Results obtained by all four methods yield an effect estimate for initiating MTX at month six that is protective, however all fail to reach statistical significance. Proximal estimates are substantially larger than the DR estimator. Interestingly, the proximal effect estimates for 12 months of MTX therapy are roughly double  estimates for MTX therapy initiated at month 6. This result suggests that an MSMM encoding a cumulative treatment effect might be appropriate for these data.

We therefore also estimated the cumulative treatment effect MSMM
\begin{equation}\label{eq:nonsaturatedreal}
    \E\{Y_{\overline a(1)}\} = \beta_0 + \beta_1 \{a(0) + a(1)\},
\end{equation}
with $\beta_1/2$ encoding the causal effect of an additional six month since MTX therapy initiation. 

We estimated $\beta_1$ using the same estimators as above.  Results are also summarized in Table \ref{tab:realresults}. Results obtained by fitting \eqref{eq:nonsaturatedreal} are similar to those from \eqref{eq:saturatedrealtwo}. All three proximal estimators indicate a significant protective effect of MTX against disease progression over the course of the first year of follow-up. The DR estimator based on SRA again gives a weaker and nonsignificant protective effect of MTX. As noted in the previous model the point estimates of an additional six months on MTX therapy in the cumulative model are indeed aligned with corresponding estimates from the saturated model. 

Our analysis reinforces our understanding of the potential protective effects of MTX on disease progression, providing more compelling evidence of such protective effects than an analysis that relies strictly on SRA.

\section{Discussion}\label{sec:dis}
We have described a novel framework for the analysis of complex longitudinal studies under a marginal structural mean model subject to potential confounding bias. The approach acknowledges that in practice, measured covariates generally fail in observational settings to capture all potential confounding mechanisms and at most may be seen as proxy measurements of underlying confounding factors. Our proximal causal inference framework provides a formal potential outcome framework under which one can articulate conditions to identify causal effects from proxies in complex longitudinal studies.

There are several possible future directions for this line of research. We note that the Cox proportional hazards MSM is widely used for censored survival time endpoints under SRA, proximal identification and inference for this model is a promising area of future research. Another possible direction for future research is to develop nonparametric proximal methods analogous to \cite{ghassami2022minimax}, however, as previously mentioned, this may be particularly challenging due to the curse of dimensionality. 

\section*{Acknowledgements}
Research reported in this publication was supported by the Beijing Natural Science Foundation, China (award Z190001, to Wang Miao), National Natural Science Foundation of China (award 12071015, to Wang Miao), and the National Institutes of Health (award R01GM139926 to Xu Shi, award R01AI27271, R01AG065276, R01GM139926, to Eric J. Tchetgen Tchetgen). Andrew Ying was awarded the David P. Byar Early Career Award under the Biometrics Section of the American Statistical Association at the 2022 Joint Statistical Meetings.

\appendix
\section{Description of the Supplementary Material}
In the supplementary material we provide a general treatment of proximal causal identification of MSMM in longitudinal studies of arbitrary follow-up. This material also includes proofs to all results given in the main text of the paper allowing for follow-up of arbitrary length. We also provide more extensive discussion of the completeness conditions used throughout the paper for the unfamiliar reader. 
An alternative characterization of the set of influence functions of $\beta_*$ which does not impose Assumption \ref{assump:ralregularitytwo} is provided. We show the unbiasedness of the estimating equations used for estimating the nuisance parameters $(h_1, h_0)$, $(q_0, q_1)$. An algorithmic summary of the steps towards constructing the proposed estimators $\widehat \beta_{\text{POR}}$, $\widehat \beta_{\text{PIPW}}$ and $\widehat \beta_{\text{PDR}}$ is given. Asymptotic distributions for $\widehat \beta_{\text{POR}}$, $\widehat \beta_{\text{PIPW}}$ and $\widehat \beta_{\text{PDR}}$ are provided. Compatibility of the confounding bridge functions with respect to the data generating process used in simulation studies is proved. Finally, supplemental simulations are summarized in this material, evaluating sensitivity of the proposed methods to violations of identifying conditions.

Code replicating numerical results including all simulations are provided on github. Here is the link: \url{https://github.com/andrewyyp/Proximal_Causal_Inference_for_Complex_Longitudinal_Studies.git}.
\bibliographystyle{chicago}

\bibliography{ref}

\begin{thebibliography}{}

\bibitem[\protect\citeauthoryear{Andrews}{Andrews}{2017}]{andrews2017examples}
Andrews, D.~W. (2017).
\newblock Examples of $\uppercase{L}^2$-complete and boundedly-complete
  distributions.
\newblock {\em Journal of Econometrics\/}~{\em 199}, 213--220.

\bibitem[\protect\citeauthoryear{Chen, Chernozhukov, Lee, and Newey}{Chen
  et~al.}{2014}]{chen2014local}
Chen, X., V.~Chernozhukov, S.~Lee, and W.~K. Newey (2014).
\newblock Local identification of nonparametric and semiparametric models.
\newblock {\em Econometrica\/}~{\em 82\/}(2), 785--809.

\bibitem[\protect\citeauthoryear{Choi, Hern{\'a}n, Seeger, Robins, and
  Wolfe}{Choi et~al.}{2002}]{choi2002methotrexate}
Choi, H.~K., M.~A. Hern{\'a}n, J.~D. Seeger, J.~M. Robins, and F.~Wolfe (2002).
\newblock Methotrexate and mortality in patients with rheumatoid arthritis: a
  prospective study.
\newblock {\em The Lancet\/}~{\em 359\/}(9313), 1173--1177.

\bibitem[\protect\citeauthoryear{Cui, Pu, Shi, Miao, and Tchetgen~Tchetgen}{Cui
  et~al.}{2020}]{cui2020semiparametric}
Cui, Y., H.~Pu, X.~Shi, W.~Miao, and E.~J. Tchetgen~Tchetgen (2020).
\newblock Semiparametric proximal causal inference.
\newblock {\em arXiv preprint arXiv:2011.08411\/}.

\bibitem[\protect\citeauthoryear{Cui and Tchetgen~Tchetgen}{Cui and
  Tchetgen~Tchetgen}{2021}]{cui2020semiparametric2}
Cui, Y. and E.~J. Tchetgen~Tchetgen (2021).
\newblock A semiparametric instrumental variable approach to optimal treatment
  regimes under endogeneity.
\newblock {\em Journal of the American Statistical Association\/}~{\em
  116\/}(533), 162--173.
\newblock PMID: 33994604.

\bibitem[\protect\citeauthoryear{Darolles, Fan, Florens, and Renault}{Darolles
  et~al.}{2011}]{darolles2011nonparametric}
Darolles, S., Y.~Fan, J.-P. Florens, and E.~Renault (2011).
\newblock Nonparametric instrumental regression.
\newblock {\em Econometrica\/}~{\em 79\/}(5), 1541--1565.

\bibitem[\protect\citeauthoryear{Deaner}{Deaner}{2020}]{deaner2018panel}
Deaner, B. (2020).
\newblock Proxy controls and panel data.
\newblock {\em arXiv preprint arXiv:1810.00283v6\/}.

\bibitem[\protect\citeauthoryear{D'Haultfoeuille}{D'Haultfoeuille}{2011}]{d2011completeness}
D'Haultfoeuille, X. (2011).
\newblock On the completeness condition in nonparametric instrumental problems.
\newblock {\em Econometric Theory\/}, 460--471.

\bibitem[\protect\citeauthoryear{Flanders, Klein, Darrow, Strickland, Sarnat,
  Sarnat, Waller, Winquist, and Tolbert}{Flanders
  et~al.}{2011}]{flanders2011method}
Flanders, W.~D., M.~Klein, L.~A. Darrow, M.~J. Strickland, S.~E. Sarnat, J.~A.
  Sarnat, L.~A. Waller, A.~Winquist, and P.~E. Tolbert (2011).
\newblock A method for detection of residual confounding in time-series and
  other observational studies.
\newblock {\em Epidemiology (Cambridge, Mass.)\/}~{\em 22\/}(1), 59.

\bibitem[\protect\citeauthoryear{Flanders, Strickland, and Klein}{Flanders
  et~al.}{2017}]{flanders2017new}
Flanders, W.~D., M.~J. Strickland, and M.~Klein (2017).
\newblock A new method for partial correction of residual confounding in
  time-series and other observational studies.
\newblock {\em American journal of epidemiology\/}~{\em 185\/}(10), 941--949.

\bibitem[\protect\citeauthoryear{Gagnon-Bartsch and Speed}{Gagnon-Bartsch and
  Speed}{2012}]{gagnon2012using}
Gagnon-Bartsch, J.~A. and T.~P. Speed (2012).
\newblock Using control genes to correct for unwanted variation in microarray
  data.
\newblock {\em Biostatistics\/}~{\em 13\/}(3), 539--552.

\bibitem[\protect\citeauthoryear{Ghassami, Ying, Shpitser, and
  Tchetgen~Tchetgen}{Ghassami et~al.}{2022}]{ghassami2022minimax}
Ghassami, A., A.~Ying, I.~Shpitser, and E.~J. Tchetgen~Tchetgen (2022).
\newblock Minimax kernel machine learning for a class of doubly robust
  functionals with application to proximal causal inference.
\newblock In {\em International Conference on Artificial Intelligence and
  Statistics}, pp.\  7210--7239. PMLR.

\bibitem[\protect\citeauthoryear{Hern{\'a}n, Brumback, and Robins}{Hern{\'a}n
  et~al.}{2001}]{hernan2001marginal}
Hern{\'a}n, M.~A., B.~Brumback, and J.~M. Robins (2001).
\newblock Marginal structural models to estimate the joint causal effect of
  nonrandomized treatments.
\newblock {\em Journal of the American Statistical Association\/}~{\em
  96\/}(454), 440--448.

\bibitem[\protect\citeauthoryear{Hern{\'a}n and Robins}{Hern{\'a}n and
  Robins}{2020}]{hernan2020causal}
Hern{\'a}n, M.~A. and J.~M. Robins (2020).
\newblock {\em Causal {I}nference: {W}hat {I}f}.
\newblock Boca Raton: Chapman \& Hall/CRC.

\bibitem[\protect\citeauthoryear{Hu and Shiu}{Hu and
  Shiu}{2018}]{hu2018nonparametric}
Hu, Y. and J.-L. Shiu (2018).
\newblock Nonparametric identification using instrumental variables: sufficient
  conditions for completeness.
\newblock {\em Econometric Theory\/}~{\em 34\/}(3), 659--693.

\bibitem[\protect\citeauthoryear{Kallus, Mao, and Uehara}{Kallus
  et~al.}{2021}]{kallus2021causal}
Kallus, N., X.~Mao, and M.~Uehara (2021).
\newblock Causal inference under unmeasured confounding with negative controls:
  A minimax learning approach.
\newblock {\em arXiv preprint arXiv:2103.14029\/}.

\bibitem[\protect\citeauthoryear{Kuroki and Pearl}{Kuroki and
  Pearl}{2014}]{kuroki2014measurement}
Kuroki, M. and J.~Pearl (2014).
\newblock Measurement bias and effect restoration in causal inference.
\newblock {\em Biometrika\/}~{\em 101\/}(2), 423--437.

\bibitem[\protect\citeauthoryear{Lipsitch, Tchetgen~Tchetgen, and
  Cohen}{Lipsitch et~al.}{2010}]{lipsitch2010negative}
Lipsitch, M., E.~J. Tchetgen~Tchetgen, and T.~Cohen (2010).
\newblock Negative controls: a tool for detecting confounding and bias in
  observational studies.
\newblock {\em Epidemiology (Cambridge, Mass.)\/}~{\em 21\/}(3), 383.

\bibitem[\protect\citeauthoryear{Mastouri, Zhu, Gultchin, Korba, Silva, Kusner,
  Gretton, and Muandet}{Mastouri et~al.}{2021}]{mastouri2021proximal}
Mastouri, A., Y.~Zhu, L.~Gultchin, A.~Korba, R.~Silva, M.~J. Kusner,
  A.~Gretton, and K.~Muandet (2021).
\newblock Proximal causal learning with kernels: Two-stage estimation and
  moment restriction.
\newblock {\em arXiv preprint arXiv:2105.04544\/}.

\bibitem[\protect\citeauthoryear{Miao, Geng, and Tchetgen~Tchetgen}{Miao
  et~al.}{2018}]{miao2018identifying}
Miao, W., Z.~Geng, and E.~J. Tchetgen~Tchetgen (2018).
\newblock Identifying causal effects with proxy variables of an unmeasured
  confounder.
\newblock {\em Biometrika\/}~{\em 105\/}(4), 987--993.

\bibitem[\protect\citeauthoryear{Miao, Shi, and Tchetgen~Tchetgen}{Miao
  et~al.}{2018}]{miao2018confounding}
Miao, W., X.~Shi, and E.~J. Tchetgen~Tchetgen (2018).
\newblock A confounding bridge approach for double negative control inference
  on causal effects.
\newblock {\em arXiv preprint arXiv:1808.04945\/}.

\bibitem[\protect\citeauthoryear{Michael, Cui, Lorch, and
  Tchetgen~Tchetgen}{Michael et~al.}{2020}]{michael2020instrumental}
Michael, H., Y.~Cui, S.~Lorch, and E.~J. Tchetgen~Tchetgen (2020).
\newblock Instrumental variable estimation of marginal structural mean models
  for time-varying treatment.
\newblock {\em arXiv e-prints\/}, arXiv--2004.

\bibitem[\protect\citeauthoryear{Newey and Powell}{Newey and
  Powell}{2003}]{newey2003instrumental}
Newey, W.~K. and J.~L. Powell (2003).
\newblock Instrumental variable estimation of nonparametric models.
\newblock {\em Econometrica\/}~{\em 71\/}(5), 1565--1578.

\bibitem[\protect\citeauthoryear{Robins, Hern{\'a}n, and Brumback}{Robins
  et~al.}{2000}]{robins2000marginalb}
Robins, J., M.~Hern{\'a}n, and B.~Brumback (2000).
\newblock Marginal structural models and causal inference in epidemiology.
\newblock {\em Epidemiology (Cambridge, Mass.)\/}~{\em 11\/}(5), 550--560.

\bibitem[\protect\citeauthoryear{Robins}{Robins}{1986}]{robins1986new}
Robins, J.~M. (1986).
\newblock A new approach to causal inference in mortality studies with a
  sustained exposure period—application to control of the healthy worker
  survivor effect.
\newblock {\em Mathematical Modelling\/}~{\em 7\/}(9-12), 1393--1512.

\bibitem[\protect\citeauthoryear{Robins}{Robins}{1987}]{robins1987graphical}
Robins, J.~M. (1987).
\newblock A graphical approach to the identification and estimation of causal
  parameters in mortality studies with sustained exposure periods.
\newblock {\em Journal of Chronic Diseases\/}~{\em 40}, 139S--161S.

\bibitem[\protect\citeauthoryear{Robins}{Robins}{1997}]{robins1997causal}
Robins, J.~M. (1997).
\newblock Causal inference from complex longitudinal data.
\newblock In {\em Latent variable modeling and applications to causality}, pp.\
   69--117. Springer.

\bibitem[\protect\citeauthoryear{Robins}{Robins}{1998}]{robins1997marginal}
Robins, J.~M. (1998).
\newblock Marginal structural models.
\newblock In {\em 1997 Proceedings of the Section on Bayesian Statistical
  Science}, pp.\  1--10. Alexandria, VA: American Statistical Association.

\bibitem[\protect\citeauthoryear{Robins}{Robins}{1999}]{robins1999association}
Robins, J.~M. (1999).
\newblock Association, causation, and marginal structural models.
\newblock {\em Synthese\/}, 151--179.

\bibitem[\protect\citeauthoryear{Robins}{Robins}{2000}]{robins2000marginal}
Robins, J.~M. (2000).
\newblock Marginal structural models versus structural nested models as tools
  for causal inference.
\newblock In {\em Statistical models in epidemiology, the environment, and
  clinical trials}, pp.\  95--133. Springer.

\bibitem[\protect\citeauthoryear{Shi, Miao, Nelson, and Tchetgen~Tchetgen}{Shi
  et~al.}{2020}]{shi2020multiply}
Shi, X., W.~Miao, J.~C. Nelson, and E.~J. Tchetgen~Tchetgen (2020).
\newblock Multiply robust causal inference with double-negative control
  adjustment for categorical unmeasured confounding.
\newblock {\em Journal of the Royal Statistical Society: Series B (Statistical
  Methodology)\/}~{\em 82\/}(2), 521--540.

\bibitem[\protect\citeauthoryear{Shi, Miao, and Tchetgen}{Shi
  et~al.}{2020}]{shi2020selective}
Shi, X., W.~Miao, and E.~T. Tchetgen (2020).
\newblock A selective review of negative control methods in epidemiology.
\newblock {\em Current Epidemiology Reports\/}~{\em 7\/}(4), 190--202.

\bibitem[\protect\citeauthoryear{Sofer, Richardson, Colicino, Schwartz, and
  Tchetgen~Tchetgen}{Sofer et~al.}{2016}]{sofer2016negative}
Sofer, T., D.~B. Richardson, E.~Colicino, J.~Schwartz, and E.~J.
  Tchetgen~Tchetgen (2016).
\newblock On negative outcome control of unobserved confounding as a
  generalization of difference-in-differences.
\newblock {\em Statistical science: a review journal of the Institute of
  Mathematical Statistics\/}~{\em 31\/}(3), 348.

\bibitem[\protect\citeauthoryear{Tchetgen~Tchetgen}{Tchetgen~Tchetgen}{2014}]{tchetgen2014control}
Tchetgen~Tchetgen, E.~J. (2014).
\newblock The control outcome calibration approach for causal inference with
  unobserved confounding.
\newblock {\em American journal of epidemiology\/}~{\em 179\/}(5), 633--640.

\bibitem[\protect\citeauthoryear{Tchetgen~Tchetgen, Michael, and
  Cui}{Tchetgen~Tchetgen et~al.}{2018}]{tchetgen2018marginal}
Tchetgen~Tchetgen, E.~J., H.~Michael, and Y.~Cui (2018).
\newblock Marginal structural models for time-varying endogenous treatments: A
  time-varying instrumental variable approach.
\newblock {\em arXiv preprint arXiv:1809.05422\/}.

\bibitem[\protect\citeauthoryear{Tchetgen~Tchetgen, Ying, Cui, Shi, and
  Miao}{Tchetgen~Tchetgen et~al.}{2020}]{tchetgen2020introduction}
Tchetgen~Tchetgen, E.~J., A.~Ying, Y.~Cui, X.~Shi, and W.~Miao (2020).
\newblock An introduction to proximal causal learning.
\newblock {\em arXiv preprint arXiv:2009.10982\/}.

\bibitem[\protect\citeauthoryear{Tennenholtz, Shalit, and Mannor}{Tennenholtz
  et~al.}{2020}]{tennenholtz2020off}
Tennenholtz, G., U.~Shalit, and S.~Mannor (2020).
\newblock Off-policy evaluation in partially observable environments.
\newblock In {\em Proceedings of the AAAI Conference on Artificial
  Intelligence}, Volume~34, pp.\  10276--10283.

\bibitem[\protect\citeauthoryear{Wang, Zhao, Hastie, and Owen}{Wang
  et~al.}{2017}]{wang2017confounder}
Wang, J., Q.~Zhao, T.~Hastie, and A.~B. Owen (2017).
\newblock Confounder adjustment in multiple hypothesis testing.
\newblock {\em Annals of Statistics\/}~{\em 45\/}(5), 1863.

\end{thebibliography}

\includepdf[pages=-]{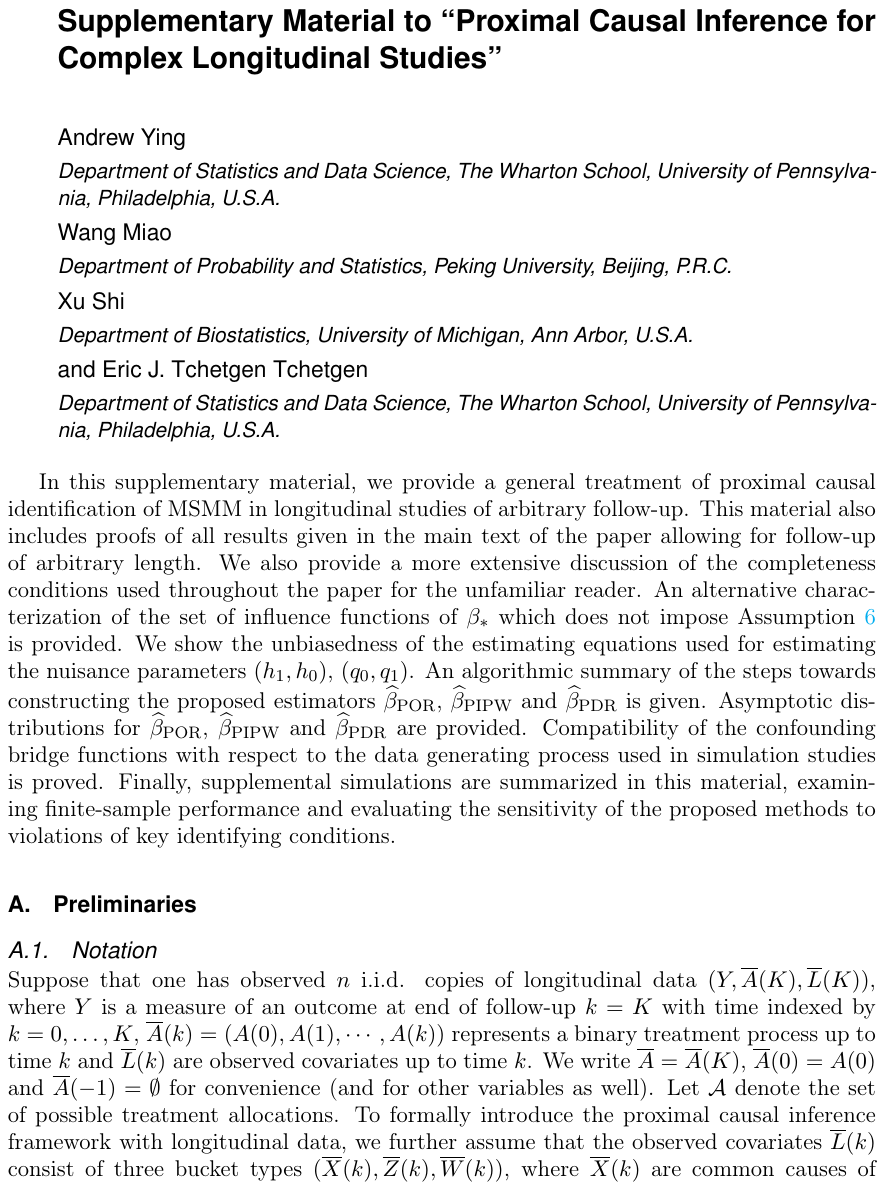}

\end{document}